\documentclass[fleqn,10pt]{wlscirep}
\usepackage[T1]{fontenc}  
 \usepackage{epsfig,epsf,psfrag} 
\usepackage{graphicx} 
\usepackage{subfigure}
\usepackage{amsmath}
\usepackage{hyperref}
\usepackage{pslatex}
\usepackage{epic,eepic} 
\usepackage{color,pstcol}
\usepackage{pstricks} 
\usepackage{fancyhdr}  \parindent0em  
\def\ua{\uparrow}
\def\da{\downarrow}
\title {Probing helicity and the topological origins of helicity via non-local Hanbury-Brown and Twiss correlations}

\author[1]{Arjun Mani}
\author[1,*]{Colin Benjamin}

\affil[1]{School of Physical Sciences, National Institute of Science Education \& Research, HBNI, Jatni-752050,\ India}

\affil[*]{colin.nano@gmail.com}


\begin{abstract}
Quantum Hall edge modes are chiral while quantum spin Hall edge modes are helical. However, unlike chiral edge modes which always occur in topological systems, quasi-helical edge modes may arise in a trivial insulator too. These trivial quasi-helical edge modes are not topologically protected and therefore need to be distinguished from helical edge modes arising due to topological reasons. Earlier conductance measurements were used to identify these helical states, in this work we report on the advantage of using the non local shot noise as a probe for the helical nature of these states as also their topological or otherwise origin and compare them with chiral quantum Hall states. We see that in similar set-ups affected by same degree of disorder and inelastic scattering, non local shot noise "HBT`` correlations can be positive for helical edge modes but are always negative for the chiral quantum Hall edge modes. Further, while trivial quasi-helical edge modes exhibit negative non-local ''HBT'' charge correlations, topological helical edge modes can show positive non-local "HBT`` charge correlation. We also study the non-local spin correlations and Fano factor for clues as regards both the distinction  between chirality/helicity as well as the topological/trivial dichotomy for helical edge modes.
\end{abstract}
\begin{document}

\flushbottom
\maketitle

\thispagestyle{empty}

\section{Introduction}

\begin{figure}
 \includegraphics[width=0.93\textwidth]{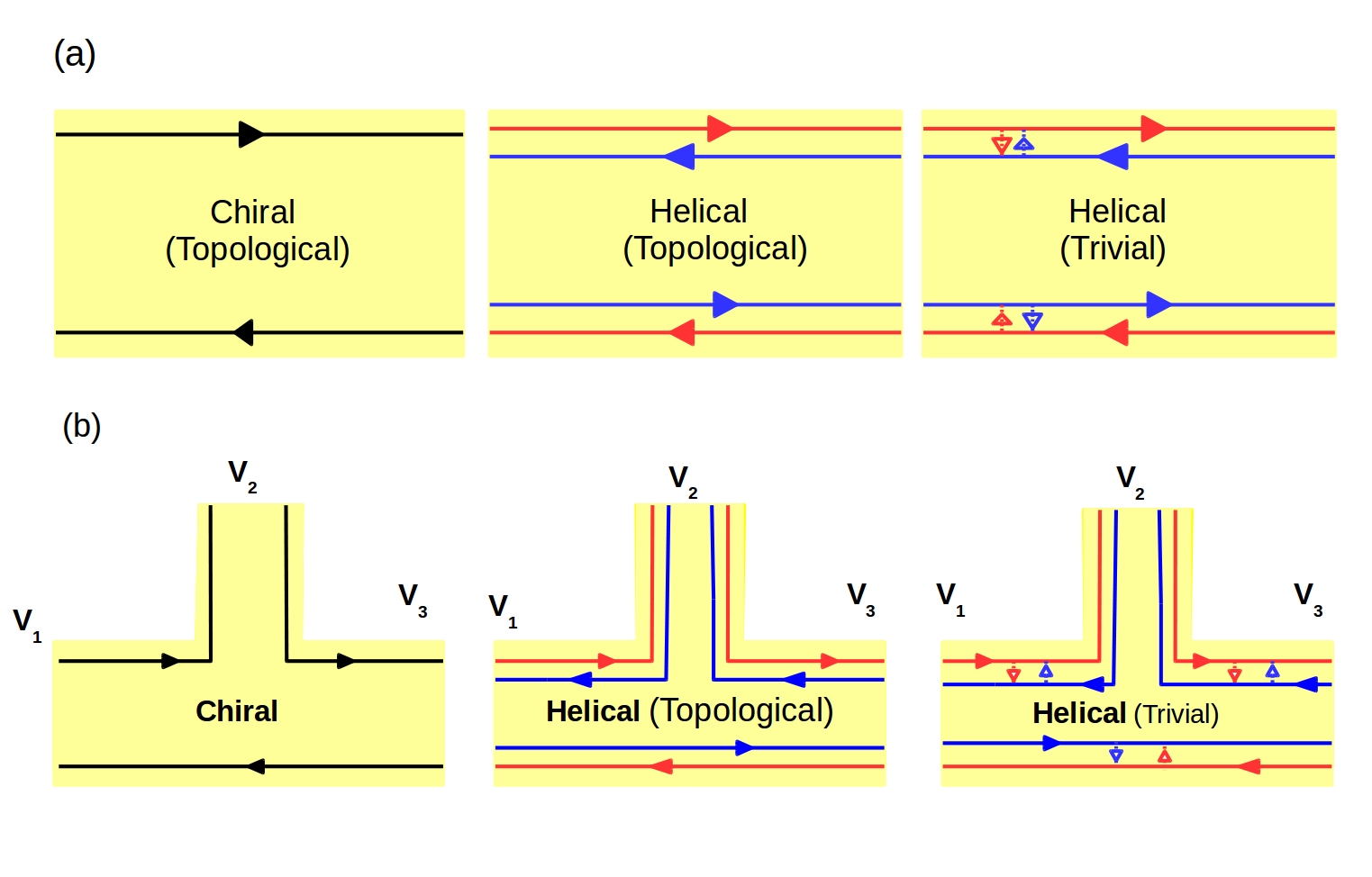}
\caption{(a) Chiral vs Helical(Topological) vs quasi-Helical (Trivial), (b) 3-terminal QH, QSH(topological) and QSH(trivial) bar}
\label{fig1}
\end{figure}

 In presence of magnetic field and at low temperatures, chiral quantum Hall (QH) edge modes appear in a 2DEG\cite{buti,datta}. In such 2D QH bars, edge modes flow in a manner (shown in  Fig.~\ref{fig1}(a), left panel) such that at the top edge electrons only move in one direction to the right. At the other, i.e., bottom edge electrons flow to the left in exactly opposite direction. So if a electron in the top edge has to change its direction, it has to scatter to the opposite edge (bottom), according to the chiral traffic rule\cite{sczhang}. At low temperatures and in samples, e.g., Mercury Telluride/Cadmium Telluride (HgTe/CdTe) heterostructure's\cite{sczhang} with strong spin-orbit coupling quantum spin Hall (QSH) edge modes appear. Herein spin of the electron is locked to its momentum. If at the top edge of the sample spin up electrons are moving in one direction (say, right) then spin down electrons are moving in the opposite direction (left). And at the bottom edge vice versa. Thus a new traffic rule comes into effect-helical traffic rule and these edge modes are therefore helical\cite{sczhang}, see Fig.~1(a) (middle panel). To scatter, an electron into the opposite direction its spin has to flip. This is prohibited by time reversal(TR) symmetry as QSH samples obey TR symmetry in contradistinction to QH samples which don't. However, it is not always that the origin of helical edge modes in QSH samples is topological, recently there have been cases\cite{marcus} where spin-momentum locked quasi-helical edge modes appear but these are not topologically protected. It has to be pointed out that the spin momentum locking among quasi-helical edge  modes does not survive non-magnetic disorder and intra-edge backscattering comes into effect. These are termed trivial quasi-helical edge modes and are shown in Fig.~1(a) (right panel). 

The reason it is necessary to probe helicity is because the QSH state is a new state of matter- it is a 2D topologically ordered phase in absence of magnetic field. The QSH edge modes due to the potent spin-orbit interaction, realize a 1D metal wherein spin is tied to the direction of motion. This unique state of matter has to be experimentally and rigorously probed such that its existence is beyond doubt. Secondly, this confusion regarding the origin of helical edge states whether its really topological and therefore protected from disorder and inelastic scattering in the sample or its due to some trivial reason and thus of non-topological origin is a current topic of interest as evidenced by the recent works in this field\cite{marcus,marcusx}.  Further, as the QSH edge modes can be used in low power information processing due to their robustness against disorder, it is very important to identify these helical edge modes and their topological origins.

There are different methods for distinguishing between helicity and chirality. The usual way to probe the existence of chiral/helical edge modes is via conductance measurements in multi terminal transport experiments\cite{Roth,Rothx}. Lets consider an elementary set-up as in Fig.~\ref{fig1}(b)(left panel)- a two terminal conductor with probes 1 and 2 as source and sink. If a third probe is added in between probe 1 and 2 as a voltage probe (Fig.~\ref{fig1}(b) (left panel)) then for QH sample one edge mode enters probe 3 from source and another edge mode goes out from it. So to maintain net current zero at probe 2 its potential is adjusted to potential of the source, and the two terminal conductance of the sample remain the same as before (without voltage probe 2). We can understand this from Landauer-Buttiker formalism\cite{buttiker,datta}. Current through voltage probe- $I_2=G_{21}(V_2-V_1)=0$ leads to $V_1=V_2$. The total conductance of the QH sample (Chiral-Topological) is then $2\frac{e^{2}}{h}$. The conductance of the QH sample does not change with addition of an extra voltage probe. But in QSH sample (topological- helical)(Fig.~\ref{fig1}(b) (middle panel)) one edge mode enters the voltage probe from source and two edge modes come out of the voltage probe. The current through voltage probe- $I_2=G_{21}(V_2-V_1)+G_{23}(V_2-V_3)=0$ leads to $I_2=G(2V_2-V_1-V_3)=0$, $\Rightarrow$ $V_2=V_1/2$, where $G_{21}=G_{23}=G$ and $V_3=0$. So its potential is adjusted to  half of the potential of the source\cite{buti-sci}. In QSH (Helical-Topological) samples the conductance is reduced by adding a voltage probe and is $\frac{3}{2}\frac{e^{2}}{h}$. Measuring the conductance with the inclusion of  a voltage probe one can differentiate between the topological helical and chiral edge modes. Now what about trivial quasi-helical edge modes these are not topologically protected and therefore these are susceptible to intra-edge scattering. At the top edge the electronic edge mode with spin-up (shown in red) has a finite probability of spin-flip scattering and reversing its path, the same thing happens for the spin-down electron (shown in blue). The small arrows in between the quasi-helical (trivial) edge modes indicates this process. The three terminal conductance for the quasi-helical trivial case then is $\frac{3}{2}\frac{e^{2}}{h} (1-f)$, where $f$ is the probability of intra-edge scattering a measure of the vulnerability of trivial quasi-helical edge modes to disorder and inelastic scattering. In clean samples where the probability of intra-edge scattering($f$) is expected to be small, relying on conductance measurement alone may not be wise. Therefore in this work we focus attention on the noise in particular the non-local HBT correlations.

Herein, we have assumed that the trivial quasi-helical edge modes in absence of any disorder are similar to the topological helical edge modes and are  spin-momentum locked. In other words- helical, up-spin edge modes at same edge have exactly opposite momentum to down-spin edge modes. There is evidence that trivial quasi-helical edge modes have spin-momentum locking\cite{marcus} but it's not fool-proof. In case there is no spin-momentum locking in the trivial phase then this case resembles a ballistic scattering state. The two terminal ballistic conductance will be double for the no spin-momentum locked trivial phase as compared to that with spin-momentum locking.  We show via a simple calculation in section 4.3.1 the validity of our assertion. However, the ballistic phase unlike the spin-momentum locked phase will be susceptible to not just spin flip scattering but also back scattering in presence of sample disorder. This may complicate the situation and again one has to take recourse to the non-local noise to resolve this situation. Thus, the topological or otherwise origin of helical edge modes needs further probe via the nonlocal HBT correlations.

The subject of this work on distinguishing topological chiral and helical edge modes and determining whether the origin of the helical edge modes is topological or not via non-local HBT correlations has not been dealt with in any previous work.  However, some works have looked at other aspects of topological helical and chiral edge modes. In two of our recent works\cite{Arjun,arjun-2} we also have explored the distinct attributes of chiral QH and helical QSH topological edge modes. Further, to the aforesaid works, few more papers \cite{hel-cooper,rashba-hel,junction,luti} have explored the topic of helical vs. chiral edge modes using superconductors\cite{hel-cooper}, with polarized STM tips\cite{luti}, with corner junctions\cite{junction} and finally exploiting the Rashba coupling\cite{rashba-hel}. All these works while relying on  different systems have a common conductance measurement which acts as the arbiter of helicity. Since in quantum spin Hall systems spin is locked to momentum, relying on just conductance measurements is risky, wherein detecting degree of spin polarization in samples exposed to disorder and spin-flip scattering will be tricky. In this paper we aim to use the Hanbury-Brown and Twiss or shot-noise correlations to probe the presence of helical edge modes and determine its origins whether topological or not. Non-local shot noise correlations on the other hand are seen to use the disorder and/or inelastic scattering present as a resource in being better able to differentiate between chirality and helicity and also between trivial and topological edge modes.

The theoretical examination of noise in QSH systems has mostly focused on the effect of electron-electron (e-e) interactions on the current-current correlations within a helical Luttinger liquid model describing the QSH state. Further these studies are in presence of a QPC in a QSH bar, as in Refs.~\cite{teo,schmidt,lee}. There are also few papers on current-current correlation studied via the scattering matrix approach or other than helical Luttinger liquid approach, see Refs.~\cite{dolci,aseev}.  In Ref.~\cite{jukka} differential noise is studied in presence of magnetic moment which is strongly dependent on frequency of the current. In our work zero frequency non-local correlations are studied for QSH systems as regards distinguishing chiral versus helical (topological) and quasi-helical (trivial) phases. The focus of the aforesaid references is not on identifying the topological origins of helical edge modes neither on the distinction between chiral and helical edge modes as is the case in our work.

However, apart from noise, various research groups around the world have made intriguing attempts at inferring helicity in edge mode transport QSH systems via the conductance-\\
A very recent proposal concerns a $\pi$ shift seen in the conductance measurement of a QSH system\cite{berry-prl},  this work relies on quantum point contacts(QPC's) which due to the Klein effect for Dirac states will be difficult to achieve experimentally  in QSH systems. This method also has an inherent weakness in that such a $\pi$ shift is only observed when backscattering is absent. This implies presence of disorder will trip this method up rendering it un-fructuous. Another interesting proposal aims to use a Hong-Ou-Mandel interferometer\cite{jonck} with QSH/QH edge modes. This work uses noise and proposes to use the dip in noise at zero power as a probe. However, this dip is shown as function of the time delay between two sources in the interferometer and its magnitude is compared for chiral and helical cases. These dips are affected by number of edge modes making the clear cut differentiation difficult. Further, no comment is made on the presence of disorder and inelastic scattering. Another work which includes disorder\cite{been-noise} and tries to distinguish between chiral and helical edge modes via a quantization of the conductance measurement obviates the weakness of Refs.\cite{berry-prl,jonck} but has an inherent weakness in that- with disorder the quantization vanishes. An interesting proposal which also uses the noise correlations\cite{z2-noise} to distinguish between chiral and helical edge modes in presence of disorder purports to be better than\cite{been-noise} but then it again would be difficult to experimentally realize with current technology because of its reliance on QPC's. Another related work concerns the amount of net  spin tunneling between edge states and this can be also used as an arbiter for helicity\cite{henrik}, however herein too effects of disorder and inelastic scattering are not dealt with, finally a related work suggests the use of noise\cite{lee} and uses a four terminal QPC to probe the helicity versus chirality dilemma, however herein too the dependence on QPC's will hamper any experimental realization. Further, the distinction between chiral and helical cases is via a difference in magnitude of the noise while a better arbiter is the sign which we will focus on in this work and will aim to surmount the challenges in the above proposals. On the question of topological helical  vs. trivial quasi-helical edge modes there have been a couple of experimental papers\cite{marcus} which have shown that quasi-helical edge modes do exist in trivial insulator but only  a single theoretical work has dealt with this problem. In Ref.\cite{gutman}, the authors propose a method to distinguish between the two which relies on the addition of two non-magnetic impurities in an other wise clean QSH sample. The occurrence of localized zero modes identifies the topological origin of the helical edge modes. Notwithstanding the complexity of the method this approach also will be hard to fashion experimentally since detecting zero modes is a non-trivial task.

  Further, while local shot noise correlations have been calculated in some recent works with QSH samples\cite{dolcini} to our knowledge this is the first work wherein both the non-local charge and spin shot noise correlations have been used as a probe of helicity and its topological origins and also to discriminate between chiral and helical edge modes. In 1950, R Hanbury Brown and R Twiss found out the diameter of radio stars via a intensity-intensity correlation experiment\cite{HBT, buttiker}. The fermionic analog of this famous experiment was realized in Ref.\cite{henny, William} for a 2DEG in the chiral QH regime. The correlations were shown to be completely anti-correlated meaning fermions in obedience to Paulli principle exclude each other. These  correlations also go by the name of shot noise which measures the correlations between fluctuations of the current\cite{blanter}. In this work we find that measuring the non-local HBT noise correlations between two disordered probes in presence of inelastic scattering via a voltage probe  one can differentiate between the chirality or helicity of the edge modes as well as the topological or trivial origin of the Helical edge modes. We observed that in the mentioned condition while the cross correlations are always negative for chiral QH case, it can be positive or negative depending on the disorder in the topological Helical QSH case. So getting a positive nonlocal correlation in these conditions identifies topological helical edge modes.  Further, the shot noise correlations can be of two types for Helical case: one the charge and the other spin. We find in this paper that while there is no distinction between charge and spin noise correlations for  topological helical edge modes, they are completely different for trivial quasi-helical edge modes enabling an effective discrimination between the topological  or trivial origins of these edge modes.

 The rest of the paper is organized as follows, first we focus on the chiral QH case and calculate the non-local "HBT'' correlations  for two disordered probes in our set-up, then proceed to case of all disordered probes. Next we add inelastic scattering to our set-up with two disordered probes and finally ending with all disordered probes with inelastic scattering. We see in all these cases non-local HBT correlations are always negative. We particularly also focus on a well known theoretical work\cite{texier} and its subsequent experimental implementation\cite{oberholzer} and show that due to the presence of quantum point contacts(QPC) in these studies, one get positive non-local correlations for chiral QH samples. In our work we deliberately remove QPC's since our focus is on obtaining positive correlations in helical QSH samples, where due to Dirac nature of the edge states experimental implementation of QPC's is difficult.  
 
 We next focus on the topological helical  QSH case, herein we distinguish between the non-local charge and spin correlations. Similar to the chiral QH case we calculate the  the non-local "HBT'' correlations  for two disordered probes in the QSH set-up, then discuss the case of all disordered probes. Like the QH case we next add inelastic scattering to the QSH set-up with two disordered probes and  finally to all disordered probes with inelastic scattering. We see that the non-local charge correlations can be positive in presence of inelastic scattering.  Next we focus on the question of distinguishing topological from trivial quasi-helical edge modes. In case of topological  helical edge modes  charge and spin HBT correlations are identical however these two differ for trivial quasi-helical case. The non-local spin HBT correlations turns completely positive for trivial quasi-helical case but the non-local charge correlations turn absolutely negative. We end the paper by focusing on the Fano factor, summarize our results in two tables, finally concluding it with a perspective on applications to other materials.
\begin{figure}[h]
 \centering {\includegraphics[width=0.75\textwidth]{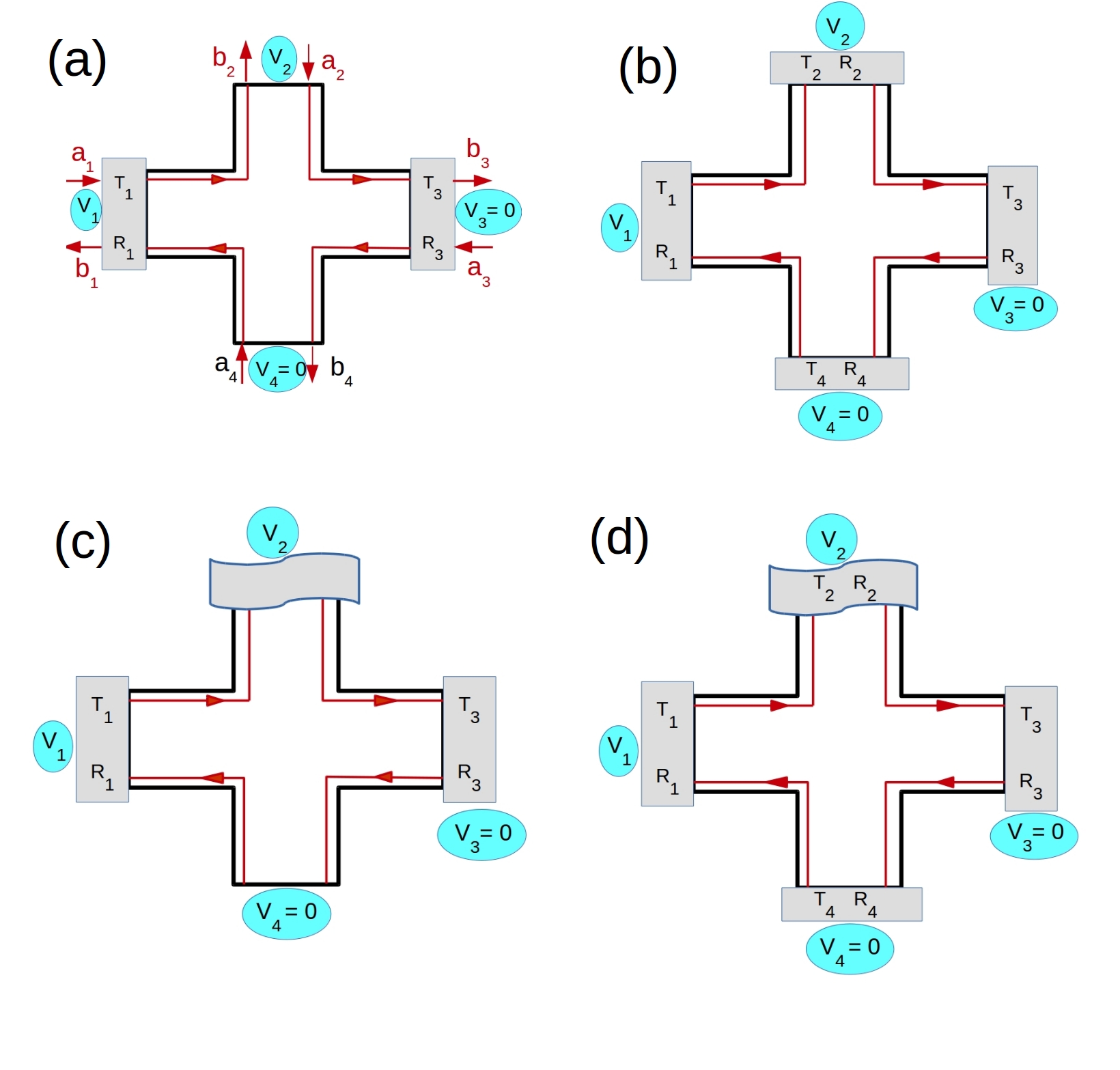}}
\caption{The QH setup, probes 3 and 4 are detectors kept at zero potential, a voltage is applied only to probe 1, (a) QH bar with two disordered probes (potential at probe 2$=0$), $a_i$'s and $b_i$'s with $i=1-4$ are the incoming and outgoing waves at the probes, (b) QH bar with all disordered probes, (c) QH bar with two disordered probes and inelastic scattering- probe 2 (curvy box) is a voltage probe (with current into it $I_{2}=0$), (d) QH bar with all disordered probes and inelastic scattering- probe 2 (curvy box) is a voltage probe (with current into it $I_{2}=0$). To avoid clutter the waves are only shown in (a). (b), (c) and (d) have exactly similar waves to and from the probes, these aren't shown explicitly.}
\label{f2}
\end{figure}
\section{Shot noise in Quantum Hall set-up}
Following the scattering matrix approach one can calculate the noise in the sample from the scattering matrix of the system as shown below\cite{texier}-
\begin{equation}\label{tex}
S_{\alpha \beta}=\frac{2e^2}{h}\int dE  \sum_{\gamma \lambda}Tr\left\{A_{\gamma \lambda}^\alpha A_{\lambda \gamma}^\beta\right\}f_\gamma(1-f_\lambda)
 \end{equation}
 wherein $A_{\gamma \lambda}^\alpha=\delta_{\alpha \gamma}\delta_{\alpha\lambda}-s_{\alpha \gamma}^\dagger s_{\alpha \lambda}$ is defined by the scattering matrix element $s_{ij}$ and $i, j$ are the indexes of the probes in the sample. In this work we detect the non-local correlations via probes \{$\alpha, \beta \rightarrow 4 , 3$ respectively. $f_\gamma$ is the Fermi-Dirac distribution at probe $\gamma$. Further, we are at zero temperature always, so $f_\gamma$ can take values 1 and 0 only.

\subsection{Quantum Hall set-up with two disordered probes} 
The two probe disordered case for QH is depicted in Fig. \ref{f2}(a). The scattering matrix relating the incoming wave to the outgoing one is given as follows:
\begin{equation}\label{QH2}
\left( \begin{array}{c} b_1\\b_2\\b_3\\b_4\end{array} \right)=s\left( \begin{array}{c}a_1\\a_2\\a_3\\a_4\end{array} \right), \mbox{with } s = \left( \begin{array}{cccc}
    r_1 & 0& 0 & t_1 \\
    -t_1  &0 & 0 & r_1 \\ 0  & -t_3 & r_3 & 0\\ 0  & r_3 & t_3 & 0 \\\end{array} \right),
\end{equation}

$r_i$ and $t_i$ represent the reflection and transmission amplitudes at contact $i$. In  Fig. \ref{f2}(a), M, the no. of edge modes is one for clarity. We can check from the scattering matrix- the unitary relation $s^\dagger s=ss^\dagger=I$ , $I$ being identity matrix, which is the necessary condition for the conservation of current.  The equations required to satisfy the scattering matrix to be unitary are $|t_i|^2+|r_i|^2=1$, where $i= $ probe index. Contact 2 is a current probe with $V_2=0$, and other potentials $V_1=V$, $V_3=V_4=0$. Thus all four contacts are basically current probes. Further, $f_1=1$ and $f_2=f_3=f_4=0$ (for $0<E<eV_1$) at zero temperature, where $E$ is the electronic energy and $f_i=$ Fermi-Dirac distribution at contact $i$ which basically depends on the potential of that contact. From Eq.~\eqref{tex} one can calculate the nonlocal correlation between probes 3 and 4 as-
\begin{eqnarray*}
S_{43}&=&\frac{2e^2}{h} \int dE \left[A_{12}^4A_{21}^3 f_1(1-f_2)+A_{13}^4A_{31}^3 f_1(1-f_3) +A_{14}^4A_{41}^3 f_1(1-f_4)\right]\\
&=&\frac{2e^2}{h}\left[e(V_1-V_2)A_{12}^4A_{21}^3+e(V_1-V_3)A_{13}^4A_{31}^3+e(V_1-V_4)A_{14}^4A_{41}^3\right]\\
&=&\frac{2e^2}{h}|eV_1|\left[s^{\dagger}_{41}s_{42}s^{\dagger}_{32}s_{31}+s^{\dagger}_{41}s_{43}s^{\dagger}_{33}s_{31}+s^{\dagger}_{41}s_{44}s^{\dagger}_{34}s_{31} \right]\\
&=&0
\end{eqnarray*}
herein $s_{ij}$ are the elements of scattering matrix(Eq.~\eqref{QH2}). Thus, from above equation, nonlocal correlation $S_{43}$ vanishes for case of two disordered probes. 
\subsection{Quantum Hall set-up with all disordered probes} 
The all probe disorder case for QH is depicted in Fig.~\ref{f2}(b). The scattering matrix relating the incoming to the outgoing wave is given as follows:
\begin{equation}\label{QHall}
s = \frac{1}{a}\left( \begin{array}{cccc}
    r_1-r_2r_3r_4 & -t_1t_2r_3r_4& -t_1t_3r_4 & -t_1t_4 \\
    -t_1t_2  &r_2-r_1r_3r_4 & -t_2t_3r_1r_4 & -t_2t_4r_1 \\ -t_1t_3r_2  & -t_2t_3 & r_3-r_1r_2r_4 & -t_3t_4r_1r_2\\ -t_1t_4r_3r_2  & -t_2t_4r_3 & -t_3t_4 & r_4-r_1r_2r_3 \\\end{array} \right),
\end{equation}
herein the term  $a=1-r_1r_2r_3r_4$ in the denominator arises because of the multiple reflections from the disordered probes\cite{Arjun}. This matrix satisfies the unitarity relation $s^\dagger s=ss^\dagger=I$. Here too the potentials are identical to two probe disorder  case- $V_1=V$, $V_2=V_3=V_4=0$. Thus, $f_1=1$ and $f_2=f_3=f_4=0$ (for $0<E<eV_1$) at zero temperature. We can calculate the nonlocal HBT correlation from Eq.~\eqref{tex} as shown below- 
\begin{eqnarray*}
S_{43}&=&\frac{2e^2}{h} \int dE \left[A_{12}^4A_{21}^3 f_1(1-f_2)+A_{13}^4A_{31}^3 f_1(1-f_3)+A_{14}^4A_{41}^3 f_1(1-f_4)\right]\\
&=&\frac{2e^2}{h}\left[e(V_1-V_2)A_{12}^4A_{21}^3+e(V_1-V_3)A_{13}^4A_{31}^3+e(V_1-V_4)A_{14}^4A_{41}^3 \right]\\
      &=&\frac{2e^2}{h}|eV_1|\left[s^{\dagger}_{41}s_{42}s^{\dagger}_{32}s_{31}+s^{\dagger}_{41}s_{43}s^{\dagger}_{33}s_{31}+s^{\dagger}_{41}s_{44}s^{\dagger}_{34}s_{31} \right]\\
&=&-\frac{2e^2}{h}|eV|\frac{T_1^2T_3T_4R_2^2R_3}{a^4}
\end{eqnarray*}
wherein we have used the unitarity or conservation of probability condition $|r_i|^2+|t_i|^2=R_i+T_i=1$. Here the correlation  depends on the disorder at probes 2 and 3 which explains why the correlation for two disordered probe (contacts 1 and 3) case is zero. The nonlocal HBT correlation is negative which is the property of the Fermi-Dirac distribution which directly relates to the antisymmetric wave function of electrons. 

\subsection{Quantum Hall set-up with disordered probes and inelastic scattering} 
We have introduced inelastic scattering via voltage probe 2. The average current $<I_2>$ through this probe is zero. Electrons coming from the probes 1 and 3 are equilibrated to a new potential at probe 2, and their phase is randomized there. The current through any probe is defined by-$I_\alpha=\frac{1}{e}\int dE \sum_{\beta}G_{\alpha \beta}f_\beta+\delta I_\alpha$, herein the second term is due to the intrinsic fluctuation and given by Eq.~\eqref{tex} and the conductance matrix $G_{\alpha \beta}=\frac{e^2}{h}(N_\alpha \delta_{\alpha \beta}-Tr\left[s_{\alpha \beta}^\dagger s_{\alpha \beta}\right])$ with $N_\alpha=$ No. of edge modes at contact $\alpha$. We need to fix the fluctuating part of the current at probe 2, $\Delta I_2=0$. This condition $\Delta I_2=0$ affects the fluctuation of current at other probes \cite{texier} as follows- 
\begin{eqnarray}\label{te}
I_\alpha=<I_\alpha>&+&\Delta I_\alpha,\mbox{with <$I_\alpha$> the average current in contact $\alpha$}\nonumber\\
\frac{1}{e}\int dE \sum_{\beta}G_{\alpha \beta}f_\beta+\delta I_\alpha=\frac{1}{e}\int dE \sum_{\beta}G_{\alpha \beta}\bar{f_\beta}&+&\Delta I_\alpha,\nonumber\\
\frac{1}{e}\int dE \sum_{\beta}G_{\alpha \beta}(f_\beta-\bar{f_\beta})+\delta I_\alpha&=&\Delta I_\alpha,\mbox{ $\bar{f_\beta}$ is average of the Fermi-Dirac distribution function in contact $\beta$,}\nonumber\\
\frac{1}{e}G_{\alpha 2}(\mu_2-\bar{\mu_2})+\delta I_\alpha&=&\Delta I_\alpha,\\
\mbox{$\mu_2$}&,&\mbox{$\bar{\mu_2}$ being chemical potential and average chemical potential at contact 2.}\nonumber
\end{eqnarray}
putting $\alpha=2$, we get
\begin{eqnarray}
\Delta I_2&=&\frac{1}{e}G_{22}(\mu_2-\bar{\mu_2})+\delta I_2,\nonumber\\
0&=&\frac{1}{e}G_{22}(\mu_2-\bar{\mu_2})+\delta I_2,\nonumber\\
\frac{\delta I_2}{G_{22}}&=&-\frac{1}{e}(\mu_2-\bar{\mu_2})\nonumber
\end{eqnarray} 
putting this in Eq.~\eqref{te} we get-
\begin{eqnarray}
\Delta I_\alpha&=&\delta I_\alpha- \frac{G_{\alpha 2}}{G_{22}}\delta I_2
\end{eqnarray} 
wherein the first term is due to the intrinsic part of the fluctuation and the second term is due to the voltage fluctuation at probe 2. Thus the nonlocal HBT correlation due to the inelastic scattering is written as-
\begin{eqnarray}\label{cor}
S_{\alpha \beta}^{in}&=&<\Delta I_\alpha \Delta I_\beta>=<(\delta I_\alpha- \frac{G_{\alpha 2}}{G_{22}}\delta I_2)(\delta I_\beta- \frac{G_{\beta 2}}{G_{22}}\delta I_2)>\nonumber\\
&=&<(\delta I_\alpha \delta I_\beta-\frac{G_{\alpha 2}}{G_{22}}\delta I_\beta \delta I_2-\frac{G_{\beta 2}}{G_{22}}\delta I_\alpha \delta I_2+\frac{G_{\alpha 2}G_{\beta 2}}{G_{22}^2}\delta I_2\delta I_2)>\nonumber\\
&=&S_{\alpha \beta}-\frac{G_{\alpha 2}}{G_{22}}S_{\beta 2}-\frac{G_{\beta 2}}{G_{22}}S_{\alpha 2}+\frac{G_{\alpha 2}G_{\beta 2}}{G_{22}^2}S_{22}
\end{eqnarray} 
In our case $\alpha =4$ and $\beta =3$.

\subsubsection{Quantum Hall set-up with two disordered probes and inelastic scattering}
 Two contacts are considered to be disordered for this case. The scattering matrix relating the incoming to the outgoing wave is given as in Eq.~\eqref{QH2}.
 Here, we have considered source $V_1=V$, and $V_3=V_4=0$ are detectors. As contact 2 is the voltage probe, 
\begin{eqnarray*}
I_2=G_{21}(V_2-V_1)+G_{24}(V_2-V_4)&=&\frac{2e^2}{h}[T_1(V_2-V_1)+R_1(V_2-V_4)]=\frac{2e^2}{h}(V_2-T_1V_1)\\
\text{putting $I_2=0$, we get-} V_2=T_1 V_{1}.
\end{eqnarray*}
The Fermi-Dirac distribution functions at zero temperature in the probes are as follows- $f_1=1$, $f_3=0$, $f_4=0$ (for $0<E<eV_1$), $f_2=1$ (for $0<E<eV_2$) and $f_2=0$ (for $eV_2<E<eV_1$) as probes 3 and 4 are used as detectors and are at zero voltage.
Following Eq.~\eqref{tex} the non-local charge correlation between probes 3 and 4 is-
{
\begin{eqnarray*}
S_{43}&=&\frac{2e^2}{h} \int dE \left[A_{12}^4A_{21}^3 f_1(1-f_2)+A_{13}^4A_{31}^3 f_1(1-f_3)+A_{14}^4A_{41}^3 f_1(1-f_4)+A_{23}^4A_{32}^3 f_2(1-f_3)+A_{24}^4A_{42}^3 f_2(1-f_4)\right]\\
&=&\frac{2e^2}{h}\left[e(V_1-V_2)A_{12}^4A_{21}^3+e(V_1-V_3)A_{13}^4A_{31}^3+e(V_1-V_4)A_{14}^4A_{41}^3+e(V_2-V_3)A_{23}^4A_{32}^3+e(V_2-V_4)A_{24}^4A_{42}^3 \right]\\
&=&\frac{2e^2}{h}|eV_1|\left[s^{\dagger}_{41}s_{42}s^{\dagger}_{32}s_{31}+s^{\dagger}_{41}s_{43}s^{\dagger}_{33}s_{31}+s^{\dagger}_{41}s_{44}s^{\dagger}_{34}s_{31} +T_1(s^{\dagger}_{42}s_{43}s^{\dagger}_{33}s_{32}+s^{\dagger}_{42}s_{44}s^{\dagger}_{34}s_{32}) \right]\\
&=&-\frac{2e^2}{h}|eV|[T_1T_3R_3]
\end{eqnarray*} 
}
Similarly, one can calculate $S_{32}=S_{42}=0$ and $S_{22}=\frac{2e^2}{h}|eV|T_1R_1$ and the conductance $G_{42}=-R_3$, $G_{32}=-T_3$ and $G_{22}=1$. 
Following Eq.~\eqref{cor} we get the non-local correlation in presence of inelastic scattering as-
\[
S_{43}^{in}=S_{43}-\frac{G_{42}}{G_{22}}S_{32}-\frac{G_{32}}{G_{22}}S_{42}+\frac{G_{32}G_{42}}{G_{22}^2}S_{22}=-\frac{2e^2}{h}|eV|[T_1^2T_3R_3]
\]

 If there are multiple no. of edge modes then the correlation is just multiplied by the no. of edge modes and it remains always negative irrespective of the disorder or inelastic scattering for QH case. 
\subsubsection{Quantum Hall set-up with all disordered probes and inelastic scattering}
All contacts are considered to be disordered for this case. The scattering matrix relating the incoming to the outgoing wave is given as in Eq.~\eqref{QHall}. In the set up as shown in Fig.~\ref{f2}(d), only one mode is shown. We have considered $V_1=V$, and $V_3=V_4=0$. As contact 2 is the voltage probe, from Landauer-Buttiker formalism putting $I_2=0$ gives $V_2=\frac{T_1V_1}{1-R_1R_3R_4}$. The Fermi-Dirac distribution functions are in the zero temperature limit given as follows- $f_1=1$, $f_3=0$, $f_4=0$ (for $0<E<eV_1$), $f_2=1$ (for $0<E<eV_2$) and $f_2=0$ (for $eV_2<E<eV_1$). From Eq.~\eqref{tex} and \eqref{cor} one can calculate the non-local correlation $S_{43}^{in}$ in presence of inelastic scattering as-
{
\begin{eqnarray*}
S_{43}^{in}=&&\frac{T_1T_3T_4R_3}{(1-R_1R_4R_3)a^8}[R_1T_2^3(1-R_3R_4)((1+R_2)(1+R_1R_3R_4)-4\sqrt{R_1R_2R_3R_4})\\-&&2 R_1T_2^2a^2(R_2+R_3R_4-R_2 R_3R_4T_1-2\sqrt{R_1R_2R_3R_4})-a^4(1-R_1R_2(1+T_2)-R_1R_2^2R_3R_4T_1)].
\end{eqnarray*}
}
\begin{figure}
 \includegraphics[width=0.5\textwidth]{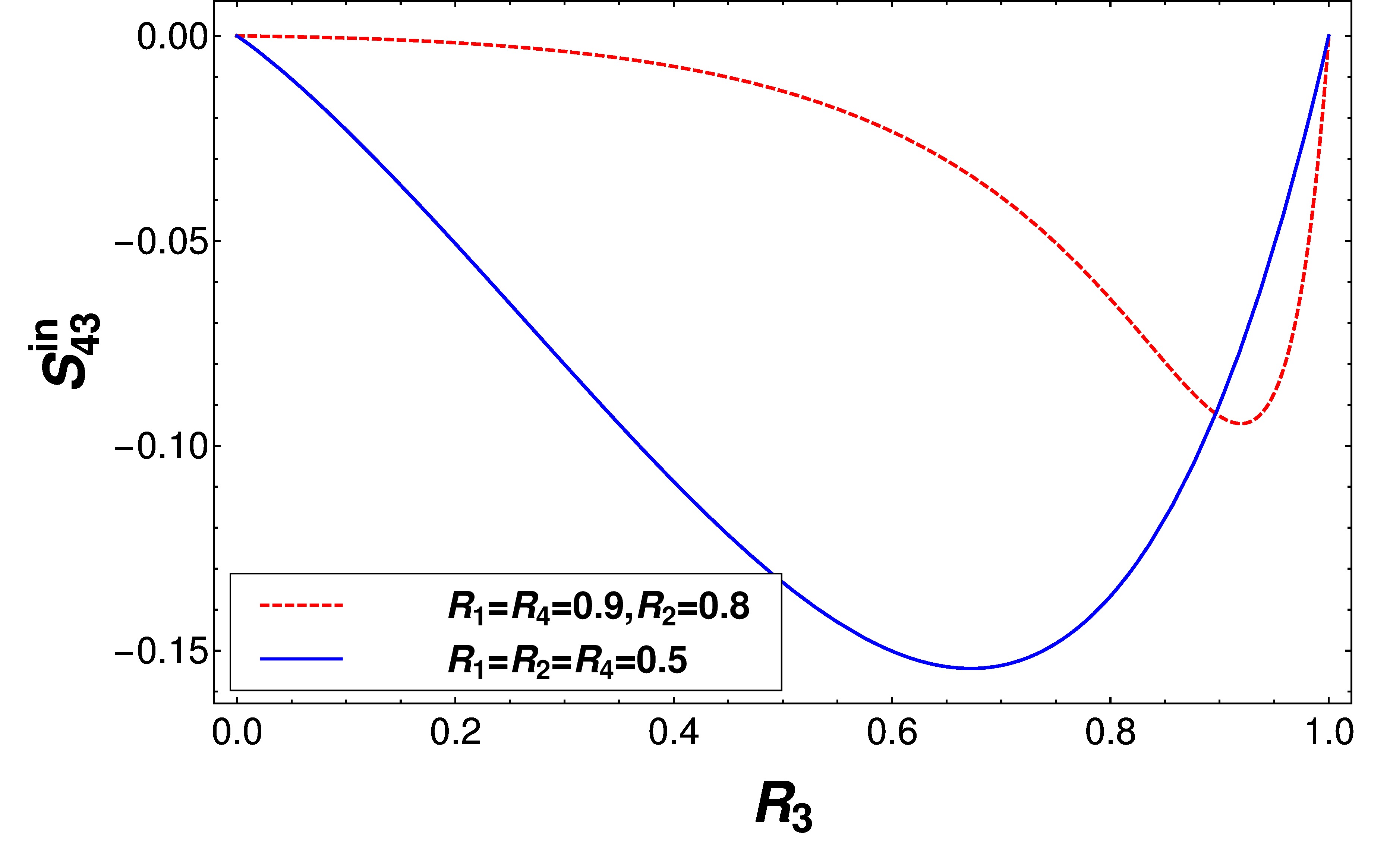}
 \caption{Non-local correlation in quantum Hall case $S_{43}^{in}$ vs $R_3$ for all disordered probes with inelastic scattering with parameters $R_1=R_2=R_4=0.5$ (solid line) and $R_1=R_4=0.9, R_2=0.8$ (dashed line).}
\label{s3}
 \end{figure}
We plot the shot noise in presence of inelastic scattering obtained from the above equation in Fig.~\ref{s3}. In Fig.~\ref{s3}(dashed line) we see as the disorder at probes 1, 2 and 4 increases the non-local correlation almost vanishes for low levels of disorder at probe 3. This is because the probe with larger disorder behaves as  closed  for the electron, meaning electron almost cannot transmit into the probe.  So it is more probable for the electron to follow a path through the probe with less disorder. This makes the electron behavior deterministic (particle like behavior) rather than probabilistic (wave like behavior), which reduces the noise correlation (almost to zero). As disorder at probe 3  increases electron path becomes more probabilistic and  negative correlations appear. One can clearly conclude that probes with same or close to the same disorder will show maximum stochastic nature in the system and will show maximum negative correlation, which is shown in Fig.~ \ref{s3}(solid line). 
 
\begin{figure}
\includegraphics[width=0.5\textwidth]{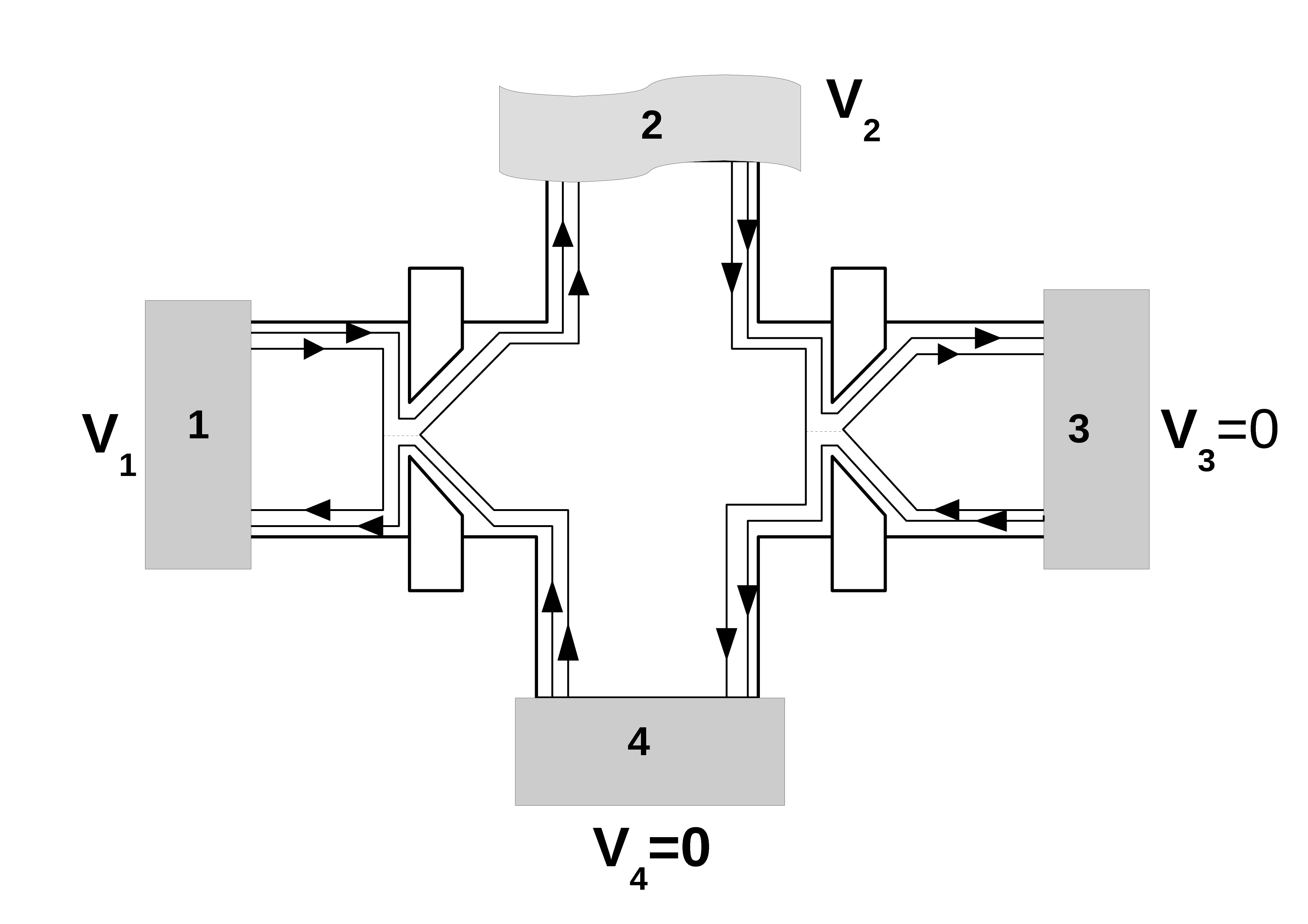}
\caption{The Texier, et. al.,/Oberholzer, et. al.,  set-up as in Refs.\cite{texier,oberholzer} to detect positive non-local HBT correlations in a quantum Hall set up. Here, probe 2 is a voltage probe ($I_{2}=0$) while probes 3 and 4 are detectors kept at zero voltage. Note that by using constrictions inside the sample and having edge modes transmitting with different probabilities one can engineer positive non-local correlations. However, in the set-ups we have in this work positive non-local correlation in quantum Hall regime are impossible.}  
\label{p4}
\end{figure}   
\subsection{Why is shot noise in our quantum Hall set-up always negative but in Texier, et. al., /Oberholzer, et. al., \cite{texier,oberholzer} set-ups it can be positive?}
Our set-up is different than Texier, et. al./Oberholzer, et. al., set-ups. They considered a constriction/QPC in their sample  which can back scatter electrons and thus creates noise within the system. In our case disorder is relegated to the probe/contact. In our case with  disorder at probes we don't have any back scattering within the sample in contrast to Texier, et. al.,/Oberholzer, et. al.,  set-ups in Fig.~\ref{p4}. Further,  in their set-ups they consider two edge modes with different transmission probabilities- one which is completely transmitted while the other is partially transmitted. However, in our case  we have identical transmission probabilities for different edge modes arising from a particular contact. Also getting a positive cross correlation in their set-up depends on the no. of edge modes in the sample but in our set-up the results are independent on the no. of edge modes. 

The  shot noise result (with inelastic scattering) derived in Ref.\cite{texier} is $S_{43}^{in}=-(e^2/h )|eV|\frac{R_3}{2}[2T_3(1+T_1)-(1+T_3)R_1T_1]$, which is positive for $T_3=0$, $S_{43}^{in}=+(e^2/h)|eV|R_1T_1/2$ for two edge modes with different transmissions. 

But if two edge modes have same transmission (lets say the two edge modes are partially transmitted with identical transmittances)  then the shot noise result is $S_{43}^{in}=-2(e^2/h )|eV|\frac{R_3}{2}[2T_3T_1-T_3R_1T_1] =-2(e^2/h )|eV|\frac{T_1T_3R_3}{2}[1+T_1]$ is completely negative as we see in our case too. The different transmittances for different edge modes arising from  a particular contact is the reason why there is a positive correlation. The experimental realization of this set-up in Ref.\cite{oberholzer} requires QPC's in order to generate different transmittances for different edge modes which for chiral QH samples maybe alright but is quite difficult for helical QSH samples, since in the latter due to Dirac nature of edge states(Klein effect) its extremely difficult to tune their transmittances via   a QPC. In this context the set-up we have which does not rely on QPC's but as we will see in next section generates positive correlations for helical edge modes becomes much more relevant for experimental implementation. Generating positive non-local correlations is the first step to generating entangled currents, which will have important applications in quantum information processing tasks.

 \section{Shot noise in quantum spin Hall set-up}
As the edge modes in QSH are spin polarized, so the non-local correlation can be calculated separately for charge as well as spin. The charge shot noise formula is given as follows- 
\begin{eqnarray}\label{ch}
S^{ch}_{\alpha \beta}&=&S_{\alpha \beta}^{\uparrow \uparrow}+S_{\alpha \beta}^{\uparrow \downarrow}+S_{\alpha \beta}^{\downarrow \uparrow}+S_{\alpha \beta}^{\downarrow \downarrow}
\end{eqnarray}
The above expression can be easily derived by extending the formalism of section II to spin with the spin shot noise formula given as- 
\begin{eqnarray}\label{sp}
S^{sp}_{\alpha \beta}&=&S_{\alpha \beta}^{\uparrow \uparrow}-S_{\alpha \beta}^{\uparrow \downarrow}-S_{\alpha \beta}^{\downarrow \uparrow}+S_{\alpha \beta}^{\downarrow \downarrow}
\end{eqnarray}

\begin{eqnarray}\label{spin}
\mbox{and, } S_{\alpha \beta}^{\sigma \sigma^\prime}&=\frac{2e^2}{h}\int dE  \sum_{\gamma \gamma^\prime} \sum_{\rho \rho^\prime =\uparrow, \downarrow}Tr\left[A_{\gamma \gamma^\prime}^{\rho \rho^\prime}(\alpha, \sigma) A_{\gamma^\prime \gamma}^{\rho^\prime \rho}(\beta, \sigma^\prime)\right]f_\gamma(1-f_{\gamma^\prime})  \nonumber
\end{eqnarray}
herein the  $\{m,n\}^{th}$ element of the Buttiker current matrix $A_{\gamma \gamma^\prime}^{\rho \rho^\prime}(\alpha, \sigma)$ is given by\cite{nikolic}-
\begin{eqnarray*}
\left[A_{\gamma \gamma^\prime}^{\rho \rho^\prime}(\alpha, \sigma)\right]_{mn}=\delta_{mn}\delta_{\gamma \alpha}\delta_{\gamma^\prime \alpha}\delta^{\sigma \rho}\delta^{\sigma \rho^\prime}-\sum_{k}\left[s_{\alpha \gamma}^{\sigma \rho\dagger}\right]_{mk}\left[s_{\alpha \gamma^\prime}^{\sigma \rho^\prime}\right]_{kn}\nonumber
\end{eqnarray*}
One can clearly see that the equations for charge and spin shot noise differ by a minus sign in front of the opposite spin correlations. This has important consequences since in presence of finite spin-flip scattering the charge and spin shot noise behave in a dis-similar manner unlike the case in absence of spin-flip wherein these are identical. 

\begin{figure}
\centering { \includegraphics[width=0.85\textwidth]{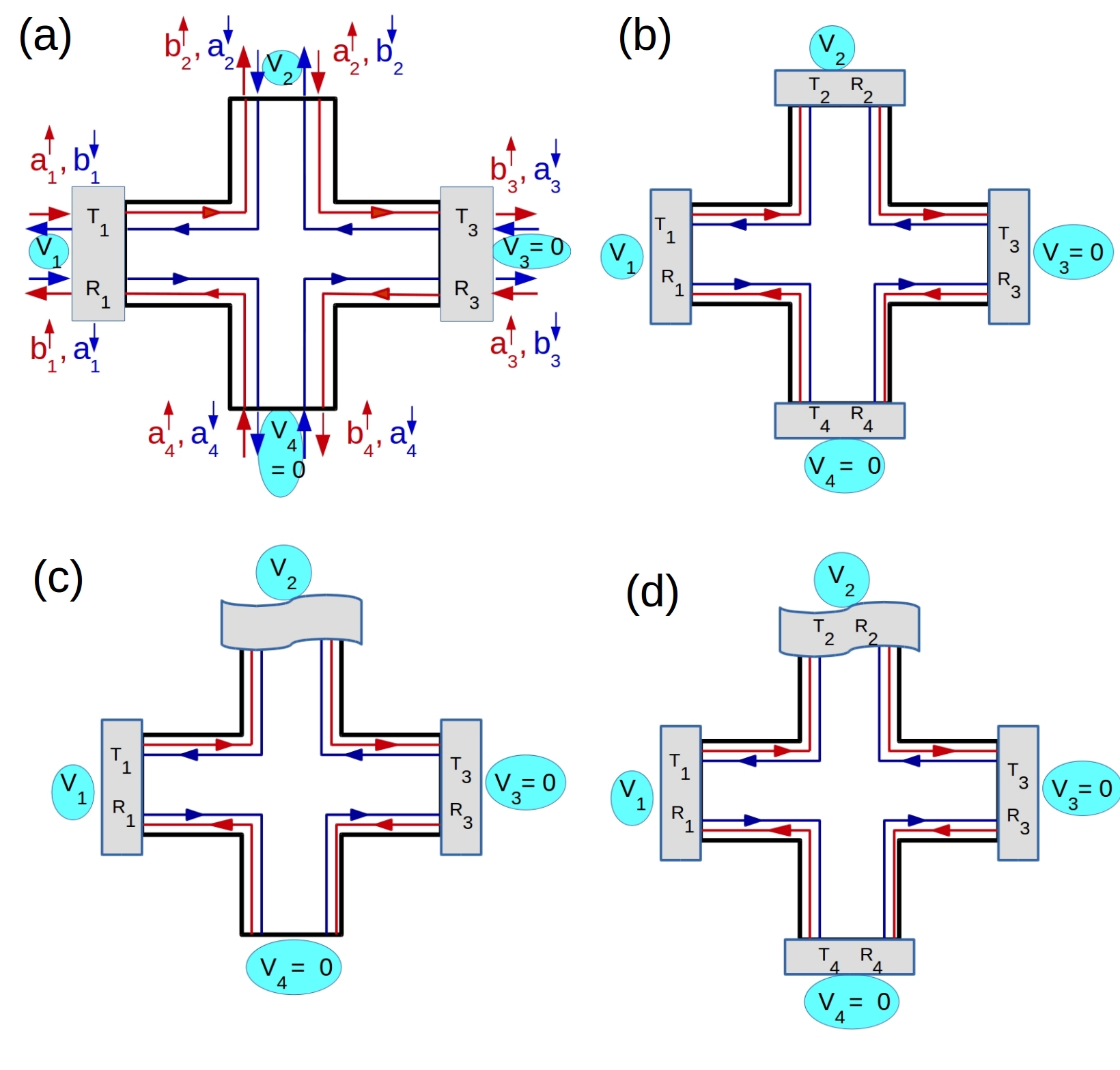}}
\vskip -0.2 in \caption{Four terminal Quantum Spin Hall bar showing QSH edge modes. These edge modes differ from their QH counterparts since these are spin polarized and helical, probes 3 and 4 are detectors kept at zero potential. Two disordered probes(a) with inelastic scattering(b) (probe 2 is a voltage probe $I_{2}=0$). $R_1=1-T_1$ and $R_3=1-T_3$, represent the reflection (transmission) probability of edge modes from and into contact $1$ and $3$ with the strength of disorder in contact 1 and 3 ranging from $0<R_{i}<1$, all disordered contacts(c)  with inelastic scattering(d). In (a) the incoming and outgoing waves into the probes are explicitly shown, these are not repeated in (b), (c) and (d) to avoid clutter. }
\label{m5}
\end{figure} 
\subsection{Quantum spin Hall set-up with two disordered probes} 
The two probe disorder case for QSH is depicted in Fig. \ref{m5}(a). The scattering matrix relating the incoming wave to the outgoing one is given as follows:

\begin{equation}\label{QSH2}
\left( \begin{array}{c} b_1^\uparrow\\b_1^\downarrow\\b_2^\uparrow\\b_2^\downarrow\\b_3^\uparrow\\b_3^\downarrow\\b_4^\uparrow\\b_4^\downarrow\end{array} \right)= s\left( \begin{array}{c}a_1^\uparrow\\a_1^\downarrow\\a_2^\uparrow\\a_2^\downarrow\\a_3^\uparrow\\a_3^\downarrow\\a_4^\uparrow\\a_4^\downarrow\end{array} \right), \mbox{with } s = \left( \begin{array}{cccccccc}
    r_1 & 0& 0 & 0& 0& 0& t_1& 0 \\
    0 & r_1& 0 & t_1& 0& 0& 0& 0 \\ -t_1 & 0& 0 & 0& 0& 0& r_1& 0\\ 0 & 0& 0 & 0& 0& t_3& 0& r_3\\0 & 0& -t_3 & 0& r_3& 0& 0& 0\\0 & 0& 0 & 0& 0& r_3& 0& -t_3\\0 & 0& r_3 & 0& t_3& 0& 0& 0\\0 & -t_1& 0 & r_1& 0& 0& 0& 0 \\\end{array} \right),
\end{equation}
with $r_i$ and $t_i$ being the reflection and transmission amplitudes at contact i. There are four probes in the case described above and the samples have two edge modes on each side- one for spin up and the other for spin down going in the opposite directions (spin-momentum locked), the scattering matrix $s$ is thus a $8 \times 8$ matrix. This matrix satisfies the unitarity relation $s^\dagger s=ss^\dagger=I$. Here the potentials are similar to QH two probe disordered  case- $V_1=V$, $V_2=V_3=V_4=0$. Further, as before at zero temperature we have the Fermi-Dirac functions as: $f_1=1$ and $f_2=f_3=f_4=0$ for ($0<E<eV_1$). From Eq.~\eqref{spin} one can calculate the nonlocal HBT correlation as-

{
 \begin{eqnarray}
S_{43}^{\uparrow \uparrow}&=&\frac{2e^2}{h} \int dE\sum_{\rho \rho^\prime =\uparrow, \downarrow} \left[A_{12}^{\rho \rho^\prime}(4,\uparrow)A_{21}^{\rho^\prime \rho}(3,\uparrow) f_1(1-f_2)+A_{13}^{\rho \rho^\prime}(4,\uparrow)A_{31}^{\rho^\prime \rho}(3,\uparrow) f_1(1-f_3)+A_{14}^{\rho \rho^\prime}(4,\uparrow)A_{41}^{\rho^\prime \rho}(3,\uparrow) f_1(1-f_4)\right]\nonumber\\
&=&\frac{2e^2}{h}\sum_{\rho \rho^\prime =\uparrow, \downarrow}\left[e(V_1-V_2)A_{12}^{\rho \rho^\prime}(4,\uparrow)A_{21}^{\rho^\prime \rho}(3,\uparrow)+ e(V_1-V_3)A_{13}^{\rho \rho^\prime}(4,\uparrow)A_{31}^{\rho^\prime \rho}(3,\uparrow)+e(V_1-V_4)A_{14}^{\rho \rho^\prime}(4,\uparrow)A_{41}^{\rho^\prime \rho}(3,\uparrow) \right]\nonumber\\
&=&\frac{2e^2}{h}|eV_1|\sum_{\rho \rho^\prime =\uparrow, \downarrow}\left[s_{41}^{\uparrow \rho\dagger}s_{42}^{\uparrow \rho^\prime}s_{32}^{\uparrow \rho^\prime\dagger}s_{31}^{\uparrow \rho}+s_{41}^{\uparrow \rho\dagger}s_{43}^{\uparrow \rho^\prime}s_{33}^{\uparrow \rho^\prime\dagger}s_{31}^{\uparrow \rho}+s_{41}^{\uparrow \rho\dagger}s_{44}^{\uparrow \rho^\prime}s_{34}^{\uparrow \rho^\prime\dagger}s_{31}^{\uparrow \rho} \right]\nonumber\\
&=&0\nonumber
\end{eqnarray}
}
Similarly one can calculate $S_{43}^{\uparrow\downarrow}=S_{43}^{\downarrow\uparrow}=S_{43}^{\downarrow\downarrow}=0$. So the sum of them $S_{43}=0$. This result is identical to QH case.

\subsection{{Quantum spin Hall set-up with all disordered probes}} The case represented in Fig.\ref{m5}(b), depicts all disordered probe case.
The scattering matrix relating the incoming  to the outgoing wave is given as follows:

\begin{equation}\label{QSHall}
{s =\frac{1 }{a} \left( \begin{array}{cccccccc}
     r_1-r_2r_3r_4 & 0& -t_1t_2r_3r_4 & 0& -t_1t_3r_4& 0& -t_1t_4& 0 \\
    0& r_1-r_2r_3r_4& 0 &-t_1t_2& 0& -t_1t_3r_2& 0& -t_1t_4r_2r_3 \\
 -t_1t_2 & 0& r_2-r_1r_3r_4 & 0& -t_2t_3r_1r_4& 0& -t_2t_4r_1& 0 \\
0&-t_1t_2r_3r_4 & 0&r_2-r_1r_3r_4  & 0& -t_2t_3& 0& -t_2t_4r_3\\
-t_1t_3r_2 & 0& -t_2t_3 & 0& r_3-r_1r_2r_4& 0& -t_3t_4r_1r_2& 0\\
0 & -t_1t_3r_4& 0 & -t_2t_3r_1r_4& 0& r_3-r_1r_2r_4& 0& -t_3t_4\\
-t_1t_4r_2r_3 & 0& -t_2t_4r_3 & 0& -t_3t_4& 0& r_4-r_1r_2r_3& 0\\
0 & -t_1t_4& 0 & -t_2t_4r_1& 0& -t_3t_4r_1r_2& 0&  r_4-r_1r_2r_3\\\end{array} \right),
}
\end{equation}

wherein $a=1-r_1r_2r_3r_4$. 
The above matrix satisfies the unitarity condition- $s^\dagger s=ss^\dagger=I$. Herein the potentials are identical to the two disordered   probes case- $V_1=V$, $V_2=V_3=V_4=0$. Here again $f_1=1$ and $f_2=f_3=f_4=0$ (for $0<E<eV_1$). From Eqs. \eqref{ch},\eqref{spin} we can calculate the non-local shot noise cross correlation-
\begin{eqnarray*}
S^{ch}_{43}=-\frac{2e^2}{h}|eV|\frac{T_1^2T_3T_4(R_2^2R_3+R_4)}{a^4}\nonumber
\end{eqnarray*}  
Thus the nonlocal charge correlation depends on the disorder at probes 2, 3 and 4 which explains why the correlation is zero for two disordered probes case (disorder at probe 1 and 3). This correlation is always negative irrespective of the magnitude of disorder.

\begin{figure*}
 \centering  
 { \includegraphics[width=.9\textwidth]{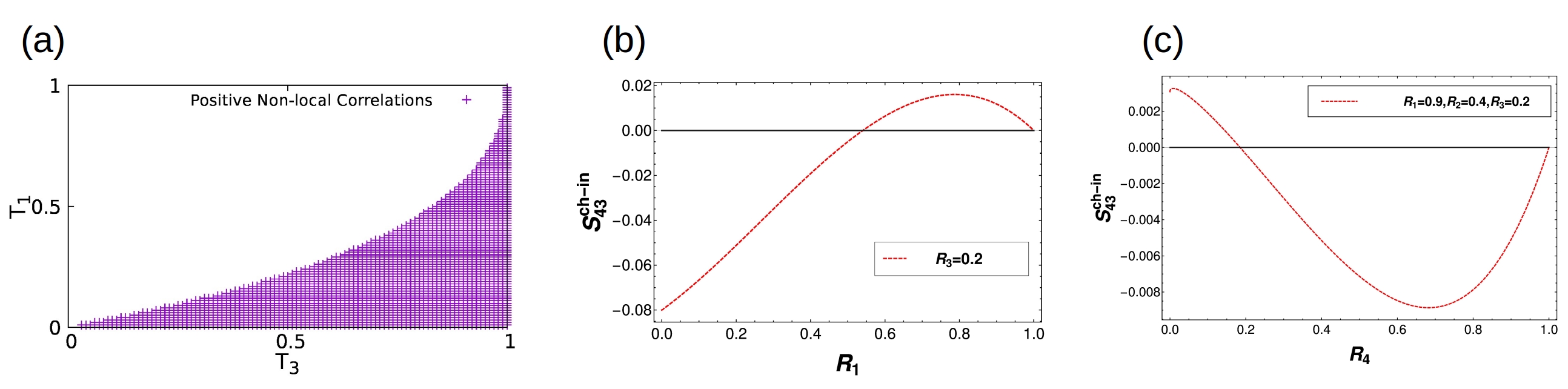}}
\caption{$S_{43}$ vs. Disorder. (a) $T_1$ vs $T_3$ for two probe disorder with inelastic scattering for QSH (Positive cross correlation), (b) correlation $S_{43}$ vs $R_1$ for two probe disorder with inelastic scattering for QSH with parameters $R_3=0.2$, (c) correlation $S_{43}$ vs $R_4$ for all probe disorder with inelastic scattering for QSH with parameters $R_1=0.9$, $R_2=0.4$ and $R_3=0.2$}
\label{n6}
\end{figure*} 

\subsection{Quantum spin Hall set-up with disorder and inelastic scattering}
To calculate the shot noise in QSH case in presence of both disorder as well as inelastic scattering we generalize the formula obtained for QH case (Eq.~7) by including spin. The non-local charge correlations for QSH case in presence of inelastic scattering then is-
\begin{eqnarray}
S_{\alpha\beta}^{ch-in}&=&S_{\alpha\beta}^{ch}-\frac{G_{\alpha2}^{ch}}{G_{22}^{ch}}S_{\beta2}^{ch}-\frac{G_{\beta2}^{ch}}{G_{22}^{ch}}S_{\alpha2}^{ch}+\frac{G_{\alpha2}^{ch}G_{\beta2}^{ch}}{{G_{22}^{ch}}^2}S_{22}^{ch}\nonumber,\\
&=&(S_{\alpha\beta}^{\uparrow\uparrow}+S_{\alpha\beta}^{\uparrow\downarrow}+S_{\alpha\beta}^{\downarrow\uparrow}+S_{\alpha\beta}^{\downarrow\downarrow})-\frac{G_{\alpha2}^{ch}}{G_{22}^{ch}}(S_{\beta2}^{\uparrow\uparrow}+S_{\beta2}^{\uparrow\downarrow}+S_{\beta2}^{\downarrow\uparrow}+S_{\beta2}^{\downarrow\downarrow})\nonumber\\
&-&\frac{G_{\beta2}^{ch}}{G_{22}^{ch}}(S_{\alpha2}^{\uparrow\uparrow}+S_{\alpha2}^{\uparrow\downarrow}+S_{\alpha2}^{\downarrow\uparrow}+S_{\alpha2}^{\downarrow\downarrow})-\frac{G_{\beta2}^{ch}G_{\alpha2}^{ch}}{{G_{22}^{ch}}^{2}}(S_{22}^{\uparrow\uparrow}+S_{22}^{\uparrow\downarrow}+S_{22}^{\downarrow\uparrow}+S_{22}^{\downarrow\downarrow})
\end{eqnarray}

Eq.~12 can be simplified by separating the individual spin components as follows:
\begin{eqnarray}
S_{\alpha\beta}^{ch-in}&=&(S_{\alpha\beta}^{\uparrow\uparrow}-\frac{G_{\alpha2}^{ch}}{G_{22}^{ch}}S_{\beta2}^{\uparrow\uparrow}-\frac{G_{\beta2}^{ch}}{G_{22}^{ch}}S_{\alpha2}^{\uparrow\uparrow}+\frac{G_{\alpha2}^{ch} G_{\beta2}^{ch}}{{G_{22}^{ch}}^2}S_{22}^{\uparrow\uparrow})+(S_{\alpha\beta}^{\uparrow\downarrow}-\frac{G_{\alpha2}^{ch}}{G_{22}^{ch}}S_{\beta2}^{\uparrow\downarrow}-\frac{G_{\beta2}^{ch}}{G_{22}^{ch}}S_{\alpha2}^{\uparrow\downarrow}+\frac{G_{\alpha2}^{ch}G_{\beta2}^{ch}}{{G_{22}^{ch}}^2}S_{22}^{\uparrow\downarrow})\nonumber\\
&&+(S_{\alpha\beta}^{\downarrow\uparrow}-\frac{G_{\alpha2}^{ch}}{G_{22}^{ch}}S_{\beta2}^{\downarrow\uparrow}-\frac{G_{\beta2}^{ch}}{G_{22}^{ch}}S_{\alpha2}^{\downarrow\uparrow}+\frac{G_{\alpha2}^{ch} G_{\beta2}^{ch}}{{G_{22}^{ch}}^2}S_{22}^{\downarrow\uparrow})+(S_{\alpha\beta}^{\downarrow\downarrow}-\frac{G_{\alpha2}^{ch}}{G_{22}^{ch}}S_{\beta2}^{\downarrow\downarrow}-\frac{G_{\beta2}^{ch}}{G_{22}^{ch}}S_{\alpha2}^{\downarrow\downarrow}+\frac{G_{\alpha2}^{ch} G_{\beta2}^{ch}}{{G_{22}^{ch}}^2}S_{22}^{\downarrow\downarrow})\nonumber\\
&=&S_{\alpha\beta}^{{\uparrow\uparrow},{in}}+S_{\alpha\beta}^{{\uparrow\downarrow},{in}}+S_{\alpha\beta}^{{\downarrow\uparrow},{in}}+S_{\alpha\beta}^{{\downarrow\downarrow},{in}}
\end{eqnarray}

In the above equation $G_{kl}^{ch}=G_{kl}^{\ua\ua}+G_{kl}^{\ua\da}+G_{kl}^{\da\ua}+G_{kl}^{\da\da}$ is the conductance summed over all the spin indices's, for example $G_{kl}^{\ua\da}$ represents the probability that  a down spin electron is transmitted as a spin up electron.
The non-local spin correlations is particular to QSH case and can be similarly, as above, written as \begin{equation}S_{\alpha\beta}^{sp-in}=S_{\alpha\beta}^{{\uparrow\uparrow},{in}}-S_{\alpha\beta}^{{\uparrow\downarrow},{in}}-S_{\alpha\beta}^{{\downarrow\uparrow},{in}}+S_{\alpha\beta}^{{\downarrow\downarrow},{in}}\end{equation} with $S_{\alpha\beta}^{ij}, i,j=\ua, \da$ defined as in Eq.~9. We proceed now by calculating the non-local charge  and spin correlations in the next sub-section and beyond.
\subsubsection{Quantum spin Hall set-up with two disordered probes and inelastic scattering} 
For the case of two disordered probes in QSH case, depicted in Fig.~ \ref{m5}(c). The scattering matrix relating the incoming to the outgoing wave is given as Eq.~\eqref{QSH2}, following Eq.~(13) we first calculate $S_{43}^{\uparrow\uparrow}$ as follows-
\begin{eqnarray*}
S_{43}^{\uparrow\uparrow}&=&\frac{2e^2}{h}\int dE \sum_{\gamma\gamma^{\prime}}\sum_{\rho\rho^{\prime}=\uparrow,\downarrow}Tr\left[A_{\gamma\gamma^{\prime}}^{\rho\rho^{\prime}}(4,\uparrow)A_{\gamma^{\prime}\gamma}^{\rho^{\prime}\rho}(3,\uparrow)\right]f_\gamma(1-f_\gamma^{\prime})\\
&=&\frac{2e^2}{h}\int dE\sum_{\gamma\gamma^{\prime}}\sum_{\rho\rho^{\prime}=\uparrow,\downarrow}Tr\left[s_{4\gamma}^{\uparrow\rho\dagger}s_{4\gamma^\prime}^{\uparrow\rho^\prime}  s_{3\gamma^\prime}^{\uparrow\rho^\prime\dagger}s_{3\gamma}^{\uparrow\rho}\right]f_\gamma(1-f_\gamma^{\prime})\\
&=&\frac{2e^2}{h}\sum_{\rho\rho^{\prime}=\uparrow,\downarrow}\left[Tr\left[s_{41}^{\uparrow\rho\dagger}s_{42}^{\uparrow\rho^\prime} s_{32}^{\uparrow\rho^\prime\dagger}s_{31}^{\uparrow\rho}\right]e(V_1-V_2)+
 Tr\left[s_{41}^{\uparrow\rho\dagger}s_{43}^{\uparrow\rho^\prime} s_{33}^{\uparrow\rho^\prime\dagger}s_{31}^{\uparrow\rho}\right]e(V_1-V_3)+ Tr\left[s_{41}^{\uparrow\rho\dagger}s_{44}^{\uparrow\rho^\prime} s_{34}^{\uparrow\rho^\prime\dagger}s_{31}^{\uparrow\rho}\right]e(V_1-V_4)+\right.\\&&\left. Tr\left[s_{42}^{\uparrow\rho\dagger}s_{43}^{\uparrow\rho^\prime} s_{33}^{\uparrow\rho^\prime\dagger}s_{32}^{\uparrow\rho}\right]e(V_2-V_3)+
 Tr\left[s_{42}^{\uparrow\rho\dagger}s_{44}^{\uparrow\rho^\prime} s_{34}^{\uparrow\rho^\prime\dagger}s_{32}^{\uparrow\rho}\right]e(V_2-V_4)\right]\\
&=&\frac{2e^2}{h}\left[0+0+0+(-T_3R_3eV_2)+0\right]\\
&=&-\frac{2e^2}{h}|eV_1|T_1T_3R_3/2
\end{eqnarray*}
Similarly from Eq.~(10) and Eq.~(13), one can calculate $S_{43}^{\uparrow\downarrow}=S_{43}^{\downarrow\uparrow}=S_{43}^{\downarrow\downarrow}=0$
then $S_{43}^{ch}=S_{43}^{\uparrow\uparrow}+S_{43}^{\uparrow\downarrow}+S_{43}^{\downarrow\uparrow}+S_{43}^{\downarrow\downarrow}=-\frac{2e^2}{h}|eV_1|T_1T_3R_3/2$.
Now to add the effect of inelastic scattering we have to calculate $S_{43}^{ch-in}$ using Eq.~(13), further  the shot noise cross-correlations $S_{32}$, $S_{42}$ and $S_{22}$ are determined following Eq.~(9). Here again we  consider $V_1=V$, and $V_3=V_4=0$. As contact 2 is the voltage probe which induces inelastic scattering, substituting $I_2=0$ gives $V_2=T_1V_1/2$. Now the Fermi-Dirac distribution functions are as follows- $f_1=1$, $f_3=0$, $f_4=0$ (for $0<E<eV_1$), $f_2=1$ (for $0<E<eV_2$) and $f_2=0$ (for $eV_2<E<eV_1$). So from Eqs.~\eqref{ch},\eqref{spin} and Eq.~(13), we calculate the non-local charge correlation in presence of inelastic scattering as-  \\
\begin{equation}
S_{43}^{ch-in}=-\frac{2e^2}{h}|eV|\left[ \frac{T_1T_3R_3}{2}-\frac{T_1T_3R_1(R_1+R_3)}{4}\right]
\end{equation}
which can be positive for a range of values of $T_1$ and $T_3$ as shown in Fig.~\ref{n6}(a). Putting $R_3=0$ we get $S_{43}^{ch-in}=\frac{2e^2}{h}|eV|[T_1R_1^2 /2]$, which is completely positive for all values of $R_1$, for $R_{3}=0.2$ we plot $S_{43}^{ch-in}$ as shown in Fig.~\ref{n6}(b). From Fig.~\ref{n6}(a) one can conclude that  small values of $T_1$ (large $R_1$) and larger values of $T_3$ (small $R_3$) help in generating positive cross correlation. In QSH case, inelastic scattering in presence of disorder induces a positive cross correlation in the system which is unexpected for electrons as they are fermions, they should show a negative cross correlation, which is the basis of the famous HBT experiment\cite{buttiker}. We can understand this in this way that QSH edge modes are spin polarized and there are spin up electrons, which after getting out of the probe 2 (voltage probe which redistributes the current) follow  one edge of the Hall bar and directly reach the contact 3 (a detector), at the same time  spin down electrons after getting out from same contact 2 follow the other edge of the Hall bar reaching contact 4 (another detector) via contact 1- and these two electrons can be correlated positively. Since these two electrons are traveling via two completely different paths and as different probes are disordered with varying degrees of  disorder these two paths  will have different transmission probabilities. But in QH case (as discussed in section II) if there are two edge modes, they do not travel via different paths (one cannot separate the paths taken by the two edge modes from voltage probe to detector) and therefore even if probes are affected with varying degrees of disorder the transmission probabilities of two edge modes will be identical. That's why positive non-local correlation is not observed in the QH set-ups as in section II even in presence of disorder and inelastic scattering. 

\subsubsection{Quantum spin Hall set-up with all disordered probes and inelastic scattering} 
The set-up for this case is depicted in Fig. \ref{m5}(b). The scattering matrix relating the incoming wave to the outgoing one is given as Eq. \eqref{QSHall}. Here we have considered $V_1=V$, and $V_3=V_4=0$. As contact 2 is the voltage probe, substituting $I_2=0$ gives $V_2=\frac{T_1V_1}{2(1-R_1R_3R_4)}$. Now the Fermi-Dirac distribution functions as usual at zero temperature and with probes 3 and 4 as detectors are as follows- $f_1=1$, $f_3=0$, $f_4=0$ (for $0<E<eV_1$), $f_2=1$ (for $0<E<eV_2$) and $f_2=0$ (for $eV_2<E<eV_1$). Thus, from Eqs.~\eqref{ch},\eqref{spin} and (14) one can calculate the non-local correlation $S_{43}^{ch-in}$, as the expressions are large we will analyze them in Fig.~\ref{n6}(c). Positive non-local correlations are obtained in this case similar to the two probe case discussed above, while inelastic scattering and the fact that up and down spin edge modes have different transmittances through the sample (due to the difference in their paths) is critical to getting positive correlations, the effect of disorder on these positive correlations is more ambiguous. Some disorder is of course necessary to have noise but other than that there is no clear cut influence of increasing/decreasing disorder on the positive correlations so obtained. 

Till now we have only considered topological Helical QSH edge modes. Now we ask the question what happens to the positive correlations so obtained if we are not sure of their topological origin. This question has become relevant recently with some papers\cite{marcus} showing that in a trivial insulator quasi-helical edge modes can also occur, of course they are without any topological protection.  In the next section we address this question.
\begin{figure}
 \centering {\includegraphics[width=.97\textwidth]{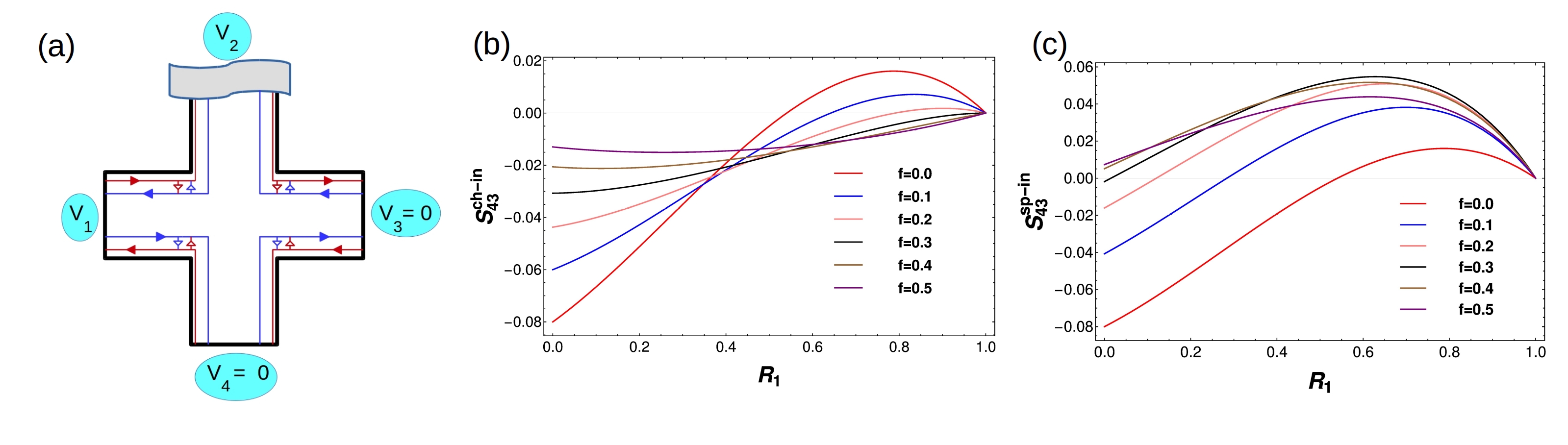}}
\caption{(a) QSH sample with trivial quasi-helical edge modes. There are two disordered probes with inelastic scattering included via voltage probe $2$, small arrowheads indicate intra edge scattering. The effect of such intra-edge scattering on positive non-local charge(b) and spin(c) correlations. Non-local charge(b) ($S_{43}^{ch-in}$ vs $R_1$) and spin(c) correlations ($S_{43}^{sp,in}$ vs $R_1$)  in a trivial quasi-helical QSH sample with two disordered probes ($R_{3}=0.2$) and inelastic scattering. Note the exactly opposite behavior to the nonlocal charge correlations. The intra edge scattering parameter: $f=0 (\mbox{topological})$ (red) and $f=0.1$ (blue), $f=0.2$(pink),$f=0.3$ (black), $f=0.4$ (brown), $f=0.5$(purple) in (b) and (c).}
\label{s7}
 \end{figure} 

\section{Topological vs. Trivial quasi-helical QSH edge modes}
We consider  trivial quasi-helical QSH edge modes as shown in Fig.~\ref{s7}(a). In Ref.\cite{marcus} the difference between trivial and topological  QSH modes is determined from the non-local Resistance. Herein we show the non-local noise (both spin as well as charge) can be very effective in determining the topological origins of QSH edge modes. Since these trivial quasi-helical edge modes are not topologically protected there is a finite probability 'f' that with disorder and inelastic scattering they will scatter from the other mode and change their direction and spin. We denote by parameter 'f'- the probability for an electron with a particular spin orientation in a trivial QSH edge mode to change its direction and spin via intra edge scattering. This intra-edge scattering is  shown in Fig.~\ref{s7}(a) by small arrows connecting two oppositely moving edge modes. Thus, 'linking' up and down spin modes due to the possibility of backscattering because of sample disorder/inelastic scattering. However, we note that in both cases trivial quasi-helical as well as topological helical QSH edge modes[see Fig.~\ref{m5}(a)], spin-momentum locking is preserved in absence of any non-magnetic disorder. An up-spin electron is backscattered as a down-spin electron moving in exactly opposite direction. 
\begin{figure}[p]
  \centering {\includegraphics[width=.98\textwidth]{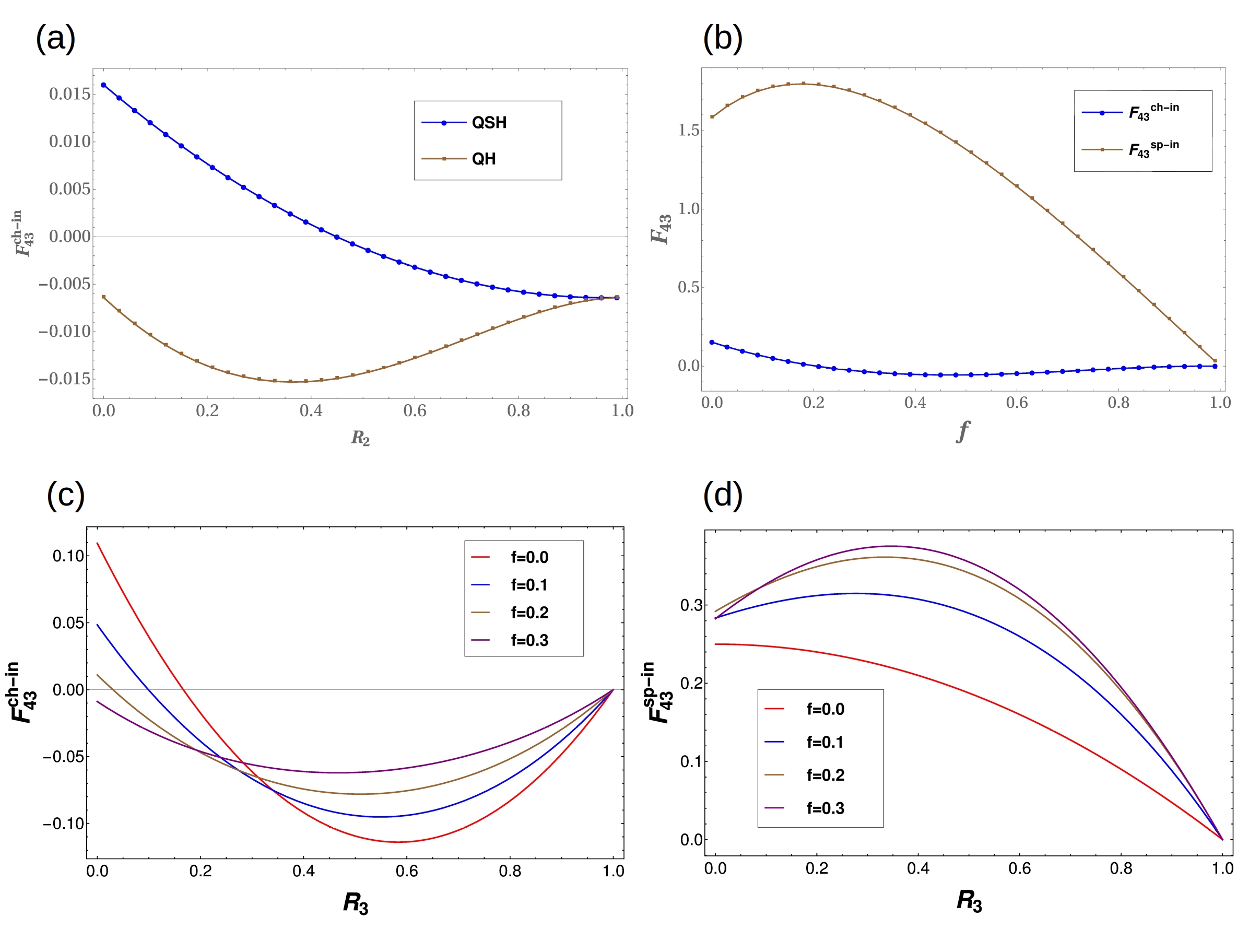}}
\caption{The effect of disorder on charge Fano factors (a) Topological QH versus Topological QSH cases, and the effect of intra-edge  scattering on Fano factors in (b) for charge and spin Fano factors in trivial QSH phase. The charge Fano factor (c) and spin Fano factor (d) in the trivial phase ($f\neq 0$) are completely distinct from topological ($f=0$) QSH phase. \\
(a) Non-local charge Fano factors in topological helical QSH and topological  chiral QH cases $F_{43}^{ch}$ vs $R_2$ for all disordered probes ($R_{4}=0, R_{3}=0.2, R_{1}=0.8$) with inelastic scattering. Intra-edge scattering probability: $f=0$. Note the sub-poissonian behavior in both cases for the charge Fano factor. (b) Non-local charge Fano factor $F_{43}^{ch}$ and spin Fano factor in trivial quasi-helical QSH sample $F_{43}^{sp}$ vs $f$ (intra-edge scattering probability) for two disordered probes ($R_{3}=0.2, R_{1}=0.8$) with inelastic scattering.  Note the super Poissonian behavior of the spin Fano factor as compared to the charge Fano factor.
(c) Non-local charge Fano factors for topological($f=0$) and trivial($f\neq 0$) QSH edge modes.  $F_{43}^{ch}$ vs $R_3$ for two disordered probes ($R_{1}=0.5$) with inelastic scattering as function of $R_3$. (d) Non-local spin Fano factors for topological($f=0$) and trivial($f\neq 0$) QSH edge modes.  $F_{43}^{sp}$ vs $R_3$ for two disordered probes ($R_{1}=0.5$) with inelastic scattering as function of $R_3$.}
\label{fano-fig}
 \end{figure} 
\subsection{ { Trivial QSH set-up with two disordered probes and inelastic scattering}} 
The case represented in Fig. \ref{s7}(a), depicts two disordered probes with inelastic scattering in a trivial QSH set-up. The scattering matrix relating the incoming to the outgoing wave is given as follows: 

\begin{equation}
{s = \left( \begin{array}{cccccccc}
    \frac{(1+f)r_1}{a_1} & \frac{-i t_1^2\sqrt{f}}{a_1}& 0 & \frac{i t_1r_1\sqrt{f(1-f)}}{a_1}&0& 0& \frac{t_1\sqrt{1-f}}{a_1}& 0 \\
    \frac{-i t_1^2\sqrt{f}}{a_1}& \frac{(1+f)r_1}{a_1}& 0 &\frac{t_1\sqrt{1-f}}{a_1}& 0&0& \frac{i t_1r_1\sqrt{f(1-f)}}{a_1} & 0 \\
\frac{-t_1\sqrt{1-f}}{a_1} & \frac{-i t_1r_1\sqrt{f(1-f)}}{a_1}& 0& \frac{i (1+R_1)\sqrt{f}}{a_1}&0& 0& \frac{r_1(1-f)}{a_1}& 0 \\
0&0&\frac{i(1+R_3)\sqrt{f}}{a_3}&0  &\frac{-i t_3r_3\sqrt{f(1-f)}}{a_3}&  \frac{-t_3\sqrt{1-f}}{a_3}&0 &\frac{r_3\sqrt{1-f}}{a_3}\\
0 & 0& \frac{t_3\sqrt{1-f}}{a_3} & 0&\frac{(1+f)r_3}{a_3}& \frac{-i t_3^2\sqrt{f}}{a_3}& 0&\frac{i t_3r_3\sqrt{f(1-f)}}{a_3} \\
0 & 0& \frac{i t_3r_3\sqrt{f(1-f)}}{a_3}  & 0&\frac{-i t_3^2\sqrt{f}}{a_3}&\frac{(1+f)r_3}{a_3} & 0&\frac{t_3\sqrt{1-f}}{a_3}  \\
0 & 0& \frac{r_3\sqrt{1-f}}{a_3} & 0&\frac{-t_3\sqrt{1-f}}{a_3} & \frac{-i t_3r_3\sqrt{f(1-f)}}{a_3}& 0&\frac{i(1+R_3)\sqrt{f}}{a_3} \\
\frac{-i t_1r_1\sqrt{f(1-f)}}{a_1} & \frac{-t_1\sqrt{1-f}}{a_1}& 0 &\frac{r_1(1-f)}{a_1} & 0& 0&\frac{i (1+R_1)\sqrt{f}}{a_1} &  0\\\end{array} \right),
}
\end{equation}

with $a_{1}=1+f r_{1}^{2}, a_{3}=1+f r_{3}^{2}$, whenever there is intra-edge scattering we introduce a $\pi/2$ phase in the scattering amplitude. 
Here too as before we have $V_1=V$, and $V_3=V_4=0$. As contact 2 is a voltage probe, putting $I_2=0$ gives- \begin{equation}                                                                                                            
V_2=\frac{(1-R_3^2f^2)T_1(1+R_1f)V_1}{2-R_1^2f(1+f)-R_3^2f(1+f)+2R_1^2R_3^2f^3}\label{eq-v2-flip}
\end{equation}
 At zero temperature, the Fermi-Dirac distribution functions are as follows: $f_1=1$, $f_3=0$, $f_4=0$ (for $0<E<eV_1$), $f_2=1$ (for $0<E<eV_2$) and $f_2=0$ (for $eV_2<E<eV_1$). From Eqs.~\eqref{ch}, \eqref{spin},(14) and (15) one can calculate the non-local charge correlation $S_{43}^{ch-in}$ as well as spin correlation $S_{43}^{sp-in}$, as the expressions are large we will analyze them via plots Fig. \ref{s7}(a) and (b).  As intra-edge scattering probability f increases to 0.25,  Fig. ~\ref{s7}(a) the correlation can be positive or negative depending on the disorder at probe 1, and as f  increases to 0.5 one can see that non-local charge correlation becomes completely negative irrespective of the disorder. Strong spin flip scattering completely destroys the positive correlation effect induced in the non-local fluctuation by the inelastic scattering in the trivial QSH sample. One can also calculate the non-local spin shot noise correlation from Eqs.~\eqref{sp},(10),(14) and (15).  This is  plotted in Fig.~\ref{s7}(c). In this case we see opposite behavior to the non-local charge correlation shown in Fig.~\ref{s7}(a). The non-local spin correlation turns completely positive with increased intra-edge scattering. Of course the non local charge and spin correlations are identical for QH case as well as for topological QSH samples. The nonlocal HBT spin correlation can thus be a good detector of trivial QSH edge modes.

\subsection{Fano Factor}
 The Fano factor, like the coefficient of variation, is a measure of the dispersion of the probability distribution of noise. It is basically the signal to noise ratio, named after Ugo Fano. Surprisingly, the noise is usually smaller than a Poisson distribution noise (in which the variance is equal to the mean value, and F=1 for Poisson distributions) and it is called sub-poissonian noise (F<1). If noise is greater than Poisson distribution then it is called super-poissonian noise. The Fano factor is defined by - $F_{ij}=\frac{S_{ij}}{2e|I|}$.
The charge Fano factor is $F_{43}^{ch}=\frac{S_{43}^{ch-in}}{2e|I_1^{ch-in}|}$, while spin Fano factor is $F_{43}^{sp}=\frac{S_{43}^{sp-in}}{2e|I_1^{sp-in}|}$. The charge current $I_1^{ch-in}=2T_1V_1-\frac{2T_1V_1f+T_1V_2(1-f)}{1-R_1f}$, and  the spin current $I_1^{sp-in}=\frac{T_1V_2(1-f)}{1+R_1f}$, where $V_2$ is defined as in Eq.~\ref{eq-v2-flip}.

We compare the charge Fano factors in the topological QH and QSH cases in Fig.\ref{fano-fig}(a) while  the charge and spin Fano factors in the trivial QSH case in Fig.\ref{fano-fig}(b). The charge Fano factor for topological QSH case changes sign while for QH case doesn't as a function of disorder. Further, in case of QSH we have two different Fano factors corresponding to charge and spin. The spin Fano factor is super-Poissonian regardless of whether the edge modes are topological or trivial while the charge Fano factor is sub-Poissonian. Thus the spin Fano factor can also be a good arbiter of the presence or absence of topological helical edge modes.  
In Figs.~8(c) the charge Fano factors are plotted as function of disorder for increasing intra edge scattering ($f$), for the topological case while Fano factor changes sign as function of disorder as intra edge scattering increases, i.e., edge modes are in trivial regime, the charge Fano factors turn more and more negative thus one can conclude that for trivial quasi-helical edge modes charge Fano factors will be negative. In Fig.~8(d) we plot the spin Fano factor, although there is no sign change but entering the trivial regime the spin Fano factor increases in magnitude for increasing intra-edge scattering $f$. The charge shot noise measured in our case is sub-Poissionian, and the charge Fano factor is well below $1/3$, which is in agreement with the experimental work of Ref.~\cite{Tikhonov} on QSH systems. We summarize the main results on distinction between chiral and helical edge modes and second between topological and trivial origins of QSH edge modes in two tables I and II. However, our story is not yet complete, the trivial phase we have assumed to have spin-momentum locked edge modes in absence of non-magnetic disorder but although there is overwhelming evidence regarding this it is not cent percent guaranteed\cite{marcus}. In the next sub-section we address this what-if regarding the trivial phase.
\begin{table}[h]
\begin{center}
\begin{tabular}{|c|c|c|c|} 
 \hline
& Chiral Edge Mode  & \multicolumn{2}{|c|}{Helical  Edge Mode}  \\ 
\hline
& Nonlocal correlations& Nonlocal charge correlations& Nonlocal spin correlations\\
\hline
Two probe disorder&0&0&0\\
\hline
All probe disorder&$-\frac{2e^2}{h}|eV|\frac{T_1^2T_3T_4R_2^2R_3}{a^4}$&$-\frac{2e^2}{h}|eV|\frac{T_1^2T_3T_4(R_2^2R_3+R_4)}{a^4}$&identical to charge\\
\hline
Two probe disorder+inelastic scatt.&$-\frac{2e^2}{h}|eV|[T_1^2T_3R_3]$&$-\frac{2e^2}{h}|eV|\left[ \frac{T_1T_3R_3}{2}-\frac{T_1T_3R_1(R_1+R_3)}{4}\right]$&identical to charge\\
\hline
All probe disorder+inelastic scatt.&negative(Fig.~\ref{s3})&Positive /Negative(Fig.~\ref{n6})&identical to charge\\
\hline
 charge Fano factor&sub-Poissonian, no sign change&sub-Poissonian, changes sign&absent \\
\hline
spin Fano factor & absent&sub-Poissonian&super-Poissonian\\
\hline
\hline
\end{tabular}
\caption{Topological Helical vs. Topological Chiral edge modes via non-local HBT correlations}
\end{center}
\end{table}

\begin{table}[h]
\begin{center}
\begin{tabular}{|c||c|c|} 
 \hline
& Topological helical  & {Trivial quasi-helical}  \\ 
\hline
Non-local Charge Noise correlations & may be positive/negative & turn completely negative  \\
\hline
Non-local Spin Noise correlations & may be positive/negative & turn completely positive \\
\hline
Charge Fano factor & changes sign & No sign change (completely negative) \\
\hline
Spin Fano factor& positive but small& positive and large\\
\hline
\hline
\end{tabular}
\caption{Topological helical vs. Trivial quasi-helical edge modes via non-local HBT correlations}
\end{center}
\end{table}
\subsection{What if the trivial quasi-helical phase does not have any spin-momentum locking, even in absence of non-magnetic disorder?}
Although its quite probable that the trivial phase in case of QSH systems is dominated by transport via quasi-helical spin-momentum locked edge modes in absence of non-magnetic disorder as shown in Fig.~\ref{s7}(a), this is by no means absolute. There might be the possibility that these might just be spin polarized ballistic modes. In fact Ref.\cite{marcus} is bit ambivalent about the exact nature of transport in the trivial phase in-spite of quite overwhelming evidence that transport in the QSH trivial phase is via spin-momentum locked edge modes in absence of any non-magnetic disorder. If this is the case then how to establish with certainty that the trivial quasi-helical phase has spin-momentum locked edge modes as discussed above or it is just ballistic modes susceptible to both back scattering as well as spin-flip scattering. To do this one can take recourse to the conductance.
\subsubsection{How good is a conductance test?}
\begin{figure}[h]
\includegraphics[width=1\textwidth]{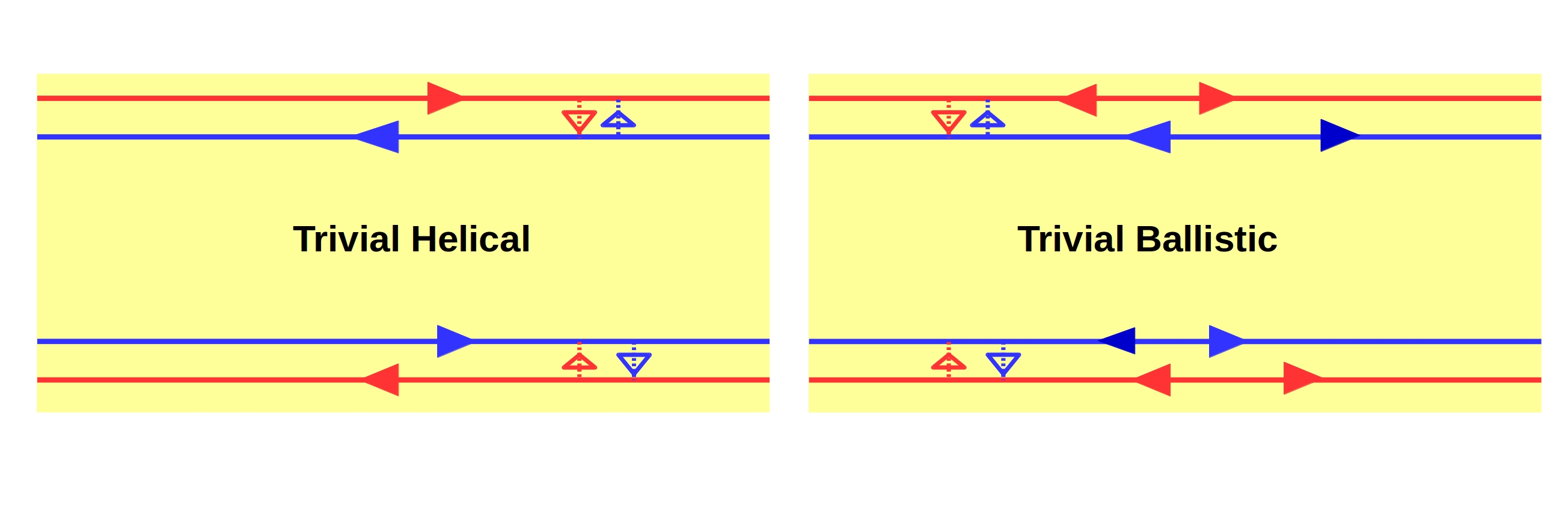}
\vspace{-0.9cm}\\
\text{\qquad\qquad\qquad\qquad(a)\qquad\qquad\qquad\qquad\qquad\qquad\qquad\qquad\qquad\qquad\qquad\qquad\qquad(b)\qquad\qquad\qquad}
\caption{(a) Spin-momentum locked trivial quasi-helical edge modes, (b) Trivial ballistic modes without spin-momentum locking. For case (b), since spin-momentum locking is broken, it is more appropriate to address these modes as ballistic although for representative comparison with the spin-momentum locking case they are shown similar to the edge modes in (a), in actuality they are anything but, the two arrows one pointing left and another pointing right on the same mode indicate backscattering while the dashed arrows linking up and down spin modes indicate spin-flip scattering.}
\label{trivial-cond}
\end{figure}

Here, we will show that the electrical conductance in two terminal/probe sample can be used to differentiate between spin-momentum locked quasi-helical trivial  phase and a trivial phase without any kind of spin-momentum locking even in absence of any non-magnetic disorder. In Ref.~\cite{marcus},  a detailed study of all the  possible Resistances ($R_{ij,kl}$) in a H shaped Hall bar in the trivial phase is done and it is shown that these resistances are quantized in absence of bulk contribution. This amounts to  the existence of edge modes in trivial phase, and needs to be distinguished from the topological helical edge modes. These quantized conductances for trivial phase only can be explained if they are made up of spin-momentum locked states (Fig.~9(a)) (with zero spin flip scattering as the sample length is below the spin-flip scattering length $l<l_\phi=4.4 \mu m$). If the spin and momentum of these states are not locked (Fig.~9(b)) then the quantized conduction will be doubled as compared to the spin-momentum locked case,  and of-course the quantized resistance is halved. We  explain this for the two terminal case in Table 3 below using the s-matrices for the quasi-helical and ballistic trivial two terminal cases as shown in Fig.~{trivial-cond}. The $4\times4$ s-matrices relating edge modes in Fig.~\ref{trivial-cond}(a) and ballistic modes in Fig.~\ref{trivial-cond}(b) are as follows-
For trivial quasi-helical edge modes on left and trivial ballistic modes on right- \begin{equation}
S_{helical}=\left( \begin{array}{cccc}
 0&-i \sqrt{f}&\sqrt{1-f}&0\\
-i\sqrt{f}&0&0&\sqrt{1-f} \\
-\sqrt{1-f}&0&0&i\sqrt{f}\\ 
0&-\sqrt{1-f}&i\sqrt{f}&0\\
\end{array} \right), S_{ballistic}=\left( \begin{array}{cccc}
 r_1 & -r_2& t_1 & t_2 \\
r_2 & r_1& t_2 & -t_1 \\
 -t_1 & -t_2& r_1 & -r_2\\ 
-t_2 & t_1& r_2 & r_1\\
 \end{array} \right)
\end{equation}
with, $t_1 = \sqrt{(1 - f)/2}$, $t_2 = \sqrt{(1 - f')/2}$, $r_1 = \sqrt{f/2}$ and $r_2 = \sqrt{f'/2} $, $f'$ denotes the backscattering probability while $f$ denotes the probability to flip. The outgoing mode basis for either s-matrix is $(b_1^\uparrow, b_1^\downarrow, b_2^\uparrow, b_2^\downarrow)$ and through $S_{ballistic/helical}$ is related to the incoming mode basis $(a_1^\uparrow, a_1^\downarrow, a_2^\uparrow, a_2^\downarrow)$. 

\begin{table}[h]
\begin{center}
\begin{tabular}{||c||c||}
\hline
Trivial quasi-helical (spin-momentum locked) phase& Trivial ballistic (spin-momentum not locked) phase\\
\hline
From Landauer-Buttiker formalism we get-&From Landauer-Buttiker formalism we get-\\
$I_i^{\sigma}=\sum_jG_{ij}^\sigma(V_i-V_j)$&$I_i^{\sigma}=\sum_jG_{ij}^{\sigma}(V_i-V_j)$\\
$I^\uparrow=G^{\uparrow}(V-0)=G^{\uparrow}V$&$I^\uparrow=G^{\uparrow}(2V-0)=2G^{\uparrow}V$\\
$I^\downarrow=G^{\downarrow}(V-0)=G^{\downarrow}V$&$I^\downarrow=G^{\downarrow}(2V-0)=2G^{\downarrow}V$\\
$G^{\uparrow}=\frac{e^2}{h}T_{21}^{\uparrow}$, with $T_{21}^{\uparrow}=T_{21}^{\uparrow\uparrow}+T_{21}^{\uparrow\downarrow}$&$G^{\uparrow}=\frac{e^2}{h}T_{21}^{\uparrow}$, with $T_{21}^{\uparrow}=T_{21}^{\uparrow\uparrow}+T_{21}^{\uparrow\downarrow}$\\
$T_{21}^{\uparrow}=(1-f)+0=(1-f)$&$T_{21}^{\uparrow}=\frac{(1-f)}{2}+\frac{(1-f')}{2}=\frac{(2-f-f')}{2}$\\
$G^{\downarrow}=\frac{e^2}{h}T_{21}^{\downarrow}=\frac{e^2}{h}(1-f)$&$G^{\downarrow}=\frac{e^2}{h}T_{21}^{\downarrow}=\frac{e^2}{h}\frac{(2-f-f')}{2}$\\
$I^{ch}=I^{\uparrow}+I^{\downarrow}=2GV$&$I^{ch}=I^{\uparrow}+I^{\downarrow}=4GV$\\
$I^{ch}=\frac{2e^2}{h}V(1-f)$&$I^{ch}=\frac{2e^2}{h}V(2-f-f')$\\
$I^{sp}=I^{\uparrow}-I^{\downarrow}=0$&$I^{sp}=I^{\uparrow}-I^{\downarrow}=0$\\

$R_{2T}=\frac{h}{2e^2 (1-f)}$&$R_{2T}=\frac{h}{2e^2(2-f-f')}$\\
\hline
\hline
\end{tabular}
\end{center}
\caption{ Charge and spin conductance in Trivial quasi-helical and Trivial ballistic phases. $f$ is the probability of spin-flip scattering, and $f'$ is the backscattering probability which is non-zero only for trivial ballistic phase. Left column has edge modes while write column has ballistic modes and therefore for $f'=0$ , $R_{2T}$ for ballistic case does not reduce to that of edge state.}
\end{table}

\begin{table}
\begin{center}
\begin{tabular}{|c|c|p{2.8 cm}|p{2.9 cm }|}
\hline
& Topological Helical Phase & \multicolumn{2}{|c|}{Trivial Phase} \\
\hline
&&quasi-Helical (spin-momentum locking)& Ballistic (no spin-momentum locking)\\
\hline
For f=0&$R_{2T}=\frac{h}{2e^2}$&$R_{2T}=\frac{h}{2e^2}$&$R_{2T}=\frac{h}{2e^2 (2-f')}$\\
\hline
For f$\neq$0&$R_{2T}=\frac{h}{2e^2}$&$R_{2T}=\frac{h}{2e^2(1-f)}$ & $R_{2T}=\frac{h}{2e^2(2-f-f')}$ \\
\hline
\end{tabular}
\caption{ Topological Helical vs. Trivial quasi-helical vs. Trivial Ballistic phase via 2T resistance}
\end{center}
\end{table}
For spin-momentum locked trivial quasi-helical phase the quantized resistance for various cases in the H shaped bar, shown in Fig.~7(a) of Ref.~\cite{marcus}, will be-
\begin{equation}
R_{14,14}=\frac{3h}{4e^2}, R_{24,14}=\frac{h}{2e^2}, R_{23,14}=\frac{h}{4e^2},  R_{23,14}=0
\end{equation}
and for the spin-momentum not locked trivial ballistic phase the quantized resistance for various case in the same H shaped bar shown in Fig.~7(a) of Ref.~\cite{marcus} will give-
\begin{equation}
R_{14,14}=\frac{3h}{8e^2}, R_{24,14}=\frac{h}{4e^2}, R_{23,14}=\frac{h}{8e^2},  R_{23,14}=0
\end{equation}
In the above equations we have considered no spin flip scattering (f=0) as well as absence of any backscattering (f'=0). In Fig.~7(c) of Ref.~\cite{marcus} they get the quantized resistance as mentioned in Eq.~18, which implies that the edge modes are spin-momentum locked. Ref.\cite{marcus} also mentions that ``scanning probe techniques demonstrate the existence of edge channels also in the inverted regime, with similarities to those measured in the trivial regime'', and `` Furthermore, the edge channel resistance per unit length is very close to earlier reports of helical edge channels." From Table 4, we see that the 2-terminal conductances for topological and trivial phase without spin-momentum locking  are different, but 2-terminal conductance for topological and trivial  phases with spin-momentum locking are same (for f=0, as the sample length is smaller than the spin-flip scattering length). This is understandable since in our systems there are no magnetic impurities so no possibility of time-reversal symmetry being broken. Thus the topological helical case wont be susceptible to spin-flip scattering due to sample disorder but trivial quasi-helical edge modes would be. Further trivial ballistic modes would be susceptible to both spin-flip as well as backscattering, since these are no longer edge modes.  A simple calculation for the 2 Terminal conductance reveals that for the trivial quasi-helical  phase it is $\frac{2e^2}{h}(1-f)$ while for the trivial ballistic phase it is $\frac{2e^2}{h}(2-f-f')$. The ballistic conductance is almost double in case of no backscattering ($f'=0$). But the situation is complicated if $f'\neq 0$.  In the latter case, one again takes recourse to the noise to determine exactly how transport is occurring in the trivial phase as shown below.

\subsubsection{Noise correlations to the rescue again}
\begin{figure}[h]
 \centering     {\includegraphics[width=.98\textwidth]{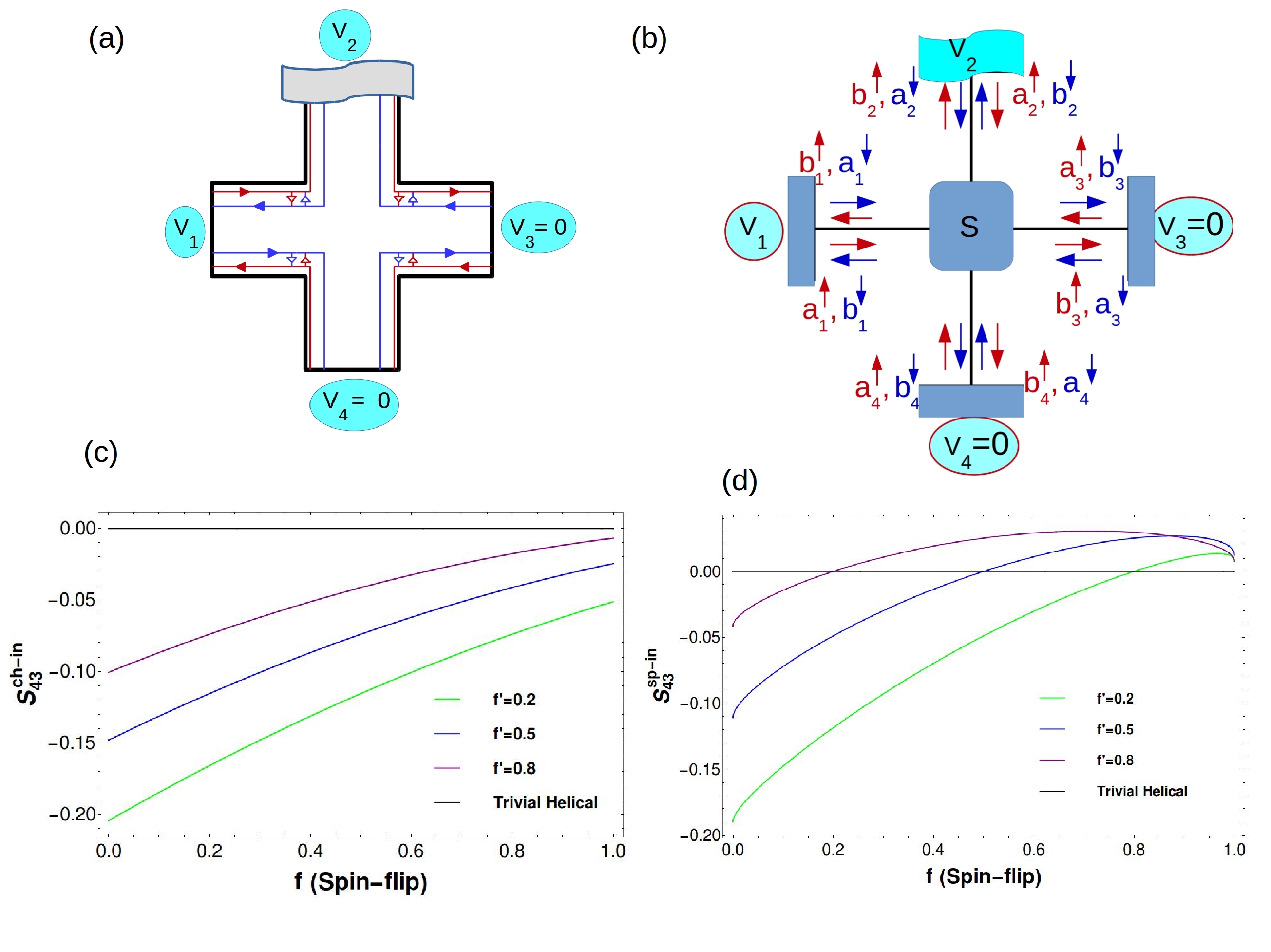}}
\caption{The set-ups to distinguish the spin-momentum locked trivial quasi-helical phase(a) from the trivial ballistic phase (b) are designed such that there are no disordered probes. Scattering happens only inside samples.  The effect of spin-flip($f$) and back-scattering ($f'$) on the non-local charge(c) and spin correlations(d). The trivial quasi-helical phase yields zero charge and spin correlations while trivial ballistic phase yields negative charge correlations and positive spin correlations. (a) The trivial quasi-helical phase: spin-momentum locked edge modes are only susceptible to spin-flip scattering. (b) The trivial ballistic phase: ballistic modes are susceptible to both spin-flip as well as backscattering. (c) Non-local charge correlations. (d)  Non-local spin correlations.}
\label{trivial}
 \end{figure}

To identify unambiguously whether the trivial phase is quasi-helical or not we consider the set-ups shown in Fig.\ref{trivial}(a,b).
 
 The set-up for the trivial quasi-helical case is depicted in Fig.~\ref{trivial}(a) while for the trivial ballistic case is depicted in Fig.~\ref{trivial}(b). The scattering matrix relating the incoming edge modes to the outgoing modes for trivial quasi-helical case is given as in Eq.~\ref{QSH-hel}. 
 \begin{equation}\label{QSH-hel}
 \left( \begin{array}{c} b_1^\uparrow\\b_1^\downarrow\\b_2^\uparrow\\b_2^\downarrow\\b_3^\uparrow\\b_3^\downarrow\\b_4^\uparrow\\b_4^\downarrow\end{array} \right)= \left( \begin{array}{cccccccc}
 0&-i \sqrt{f}&0&0&0&0&\sqrt{1-f}&0 \\
-i\sqrt{f}&0&0&\sqrt{1-f}&0&0&0&0 \\
-\sqrt{1-f}&0&0&i\sqrt{f}&0&0&0&0\\ 
0&0&i\sqrt{f}&0&0&-\sqrt{1-f}&0&0\\
0&0&\sqrt{1-f}&0&0&-i\sqrt{f}&0&0\\
0&0&0&0&-i\sqrt{f}&0&0&\sqrt{1-f}\\
0&0&0&0&-\sqrt{1-f}&0&0&i\sqrt{f}\\
0&-\sqrt{1-f}&0&0&0&0&i\sqrt{f}&0 \\\end{array} \right) \left( \begin{array}{c}a_1^\uparrow\\a_1^\downarrow\\a_2^\uparrow\\a_2^\downarrow\\a_3^\uparrow\\a_3^\downarrow\\a_4^\uparrow\\a_4^\downarrow\end{array} \right),
\end{equation}
 Since there is no disorder at the probes, we have only the spin-flip probability occurring in Eq.~\ref{QSH-hel}. Further, we have considered $V_1=V$, and $V_3=V_4=0$ and contact 2 is a voltage probe, substituting $I_2=0$ gives $V_2=V/2$. The Fermi-Dirac distribution functions as usual at zero temperature with probes 3 and 4 as detectors are as follows- $f_1=1$, $f_3=0$, $f_4=0$ (for $0<E<eV_1$), $f_2=1$ (for $0<E<eV_2$) and $f_2=0$ (for $eV_2<E<eV_1$). Thus, from Eqs.~(10) and (14) one can calculate the non-local correlation $S_{43}^{ch}=S_{42}^{ch}=0$, $S_{32}^{ch}=-\frac{2e^2}{h}eV/2f(1-f)$, and $S_{22}^{ch}=\frac{2e^2}{h}|eV| f(1-f)$. The two terminal charge charge conductances are $G_{42}=0$, $G_{32}=-(1-f)$ and $G_{22}=2(1-f)$. Putting these values in Eqs.~(12) and (14) we get $S_{43}^{ch-in}=S_{43}^{sp-in}=0$ for the trivial quasi-helical case with set-up as shown in Fig.\ref{trivial}(a).
 
 The scattering matrix relating the incoming wave to the outgoing one for trivial ballistic case is given as in Eq.~\ref{QSH2}.
\begin{equation}\label{QSH2}
 \left( \begin{array}{c} b_1^\uparrow\\b_1^\downarrow\\b_2^\uparrow\\b_2^\downarrow\\b_3^\uparrow\\b_3^\downarrow\\b_4^\uparrow\\b_4^\downarrow\end{array} \right)= \left( \begin{array}{cccccccc}
 r_1 & -r_2& t_1 & t_2& t_1& -t_2& t_1& t_2 \\
r_2 & r_1& t_2 & -t_1& t_2& t_1& t_2& -t_1 \\
 t_1 & t_2& -r_1 & r_2& -t_1& t_2& t_1& t_2\\ 
-t_2 & t_1& r_2 & r_1& t_2& t_1& -t_2& t_1\\
 t_1 & t_2& t_1 & t_2& -r_1& -r_2& -t_1& -t_2\\
-t_2 & t_1&-t_2 & t_1& r_2& -r_1& t_2& -t_1\\
 -t_1 & -t_2& t_1 & t_2& -t_1& t_2& r_1& -r_2\\
t_2 & -t_1& -t_2 & t_1& t_2& t_1& -r_2& -r_1 \\\end{array} \right) \left( \begin{array}{c}a_1^\uparrow\\a_1^\downarrow\\a_2^\uparrow\\a_2^\downarrow\\a_3^\uparrow\\a_3^\downarrow\\a_4^\uparrow\\a_4^\downarrow\end{array} \right),
\end{equation}
Here, $t_1 = \sqrt{(1 - f)/6}$, $t_2 = \sqrt{(1 - f')/6}$, $r_1 = \sqrt{f/2}$ and $r_2 = \sqrt{f'/2} $, $f'$ denotes the backscattering probability while $f$ denotes the probability to flip. 

{\begin{eqnarray}
S_{43}^{\uparrow \uparrow}&=&\frac{2e^2}{h} \int dE\sum_{\rho \rho^\prime =\uparrow, \downarrow} \left[A_{12}^{\rho \rho^\prime}(4,\uparrow)A_{21}^{\rho^\prime \rho}(3,\uparrow) f_1(1-f_2)+A_{13}^{\rho \rho^\prime}(4,\uparrow)A_{31}^{\rho^\prime \rho}(3,\uparrow) f_1(1-f_3)\right.\nonumber\\&+&\left.A_{14}^{\rho \rho^\prime}(4,\uparrow)A_{41}^{\rho^\prime \rho}(3,\uparrow) f_1(1-f_4)+A_{23}^{\rho \rho^\prime}(4,\uparrow)A_{32}^{\rho^\prime \rho}(3,\uparrow) f_2(1-f_3)+A_{24}^{\rho \rho^\prime}(4,\uparrow)A_{42}^{\rho^\prime \rho}(3,\uparrow) f_2(1-f_4)\right]\nonumber\\
&=&\frac{2e^2}{h}\sum_{\rho \rho^\prime =\uparrow, \downarrow}\left[e(V_1-V_2)A_{12}^{\rho \rho^\prime}(4,\uparrow)A_{21}^{\rho^\prime \rho}(3,\uparrow)+ e(V_1-V_3)A_{13}^{\rho \rho^\prime}(4,\uparrow)A_{31}^{\rho^\prime \rho}(3,\uparrow)+e(V_1-V_4)A_{14}^{\rho \rho^\prime}(4,\uparrow)A_{41}^{\rho^\prime \rho}(3,\uparrow)\right.\nonumber\\&+&\left.e(V_2-V_3)A_{23}^{\rho \rho^\prime}(4,\uparrow)A_{32}^{\rho^\prime \rho}(3,\uparrow)+e(V_2-V_4)A_{24}^{\rho \rho^\prime}(4,\uparrow)A_{42}^{\rho^\prime \rho}(3,\uparrow) \right]\nonumber\\
&=&\frac{2e^2}{h}|eV_1|\sum_{\rho \rho^\prime =\uparrow, \downarrow}\left[s_{41}^{\uparrow \rho\dagger}s_{42}^{\uparrow \rho^\prime}s_{32}^{\uparrow \rho^\prime\dagger}s_{31}^{\uparrow \rho}+s_{41}^{\uparrow \rho\dagger}s_{43}^{\uparrow \rho^\prime}s_{33}^{\uparrow \rho^\prime\dagger}s_{31}^{\uparrow \rho}+s_{41}^{\uparrow \rho\dagger}s_{44}^{\uparrow \rho^\prime}s_{34}^{\uparrow \rho^\prime\dagger}s_{31}^{\uparrow \rho}+s_{42}^{\uparrow \rho\dagger}s_{43}^{\uparrow \rho^\prime}s_{33}^{\uparrow \rho^\prime\dagger}s_{32}^{\uparrow \rho}+s_{42}^{\uparrow \rho\dagger}s_{44}^{\uparrow \rho^\prime}s_{34}^{\uparrow \rho^\prime\dagger}s_{32}^{\uparrow \rho} \right]\nonumber\\
&=&-\frac{2e^2}{h}\frac{|2eV_1|}{3}(t_1^2+t_2^2)^2\nonumber
\end{eqnarray}
}
 Here we consider $V_1=V$, and $V_3=V_4=0$. As contact 2 is the voltage probe, substituting $I_2=0$ gives $V_2=V/3$. Now the Fermi-Dirac distribution functions as usual at zero temperature and with probes 3 and 4 as detectors are as follows- $f_1=1$, $f_3=0$, $f_4=0$ (for $0<E<eV_1$), $f_2=1$ (for $0<E<eV_2$) and $f_2=0$ (for $eV_2<E<eV_1$). Thus, from Eqs.~(10), and (14) one can calculate the non-local charge and spin correlation as: 
\begin{eqnarray}
S_{43}^{ch-in}&=&-\frac{2e^2}{h}(2/81) |eV| (10 + 2 f^2 - 9 f' + 2 f'^2 + f (-9 + 4 f')),\nonumber\\ 
S_{43}^{sp-in}&=&-\frac{2e^2}{h}(4/81) |eV| (5 + f^2 - 6 f' + f'^2 -3 \sqrt{-(-1 + f) f} \sqrt{-(-1 + f') f'} + f (-6 + 5 f')). 
\end{eqnarray}
The non-local charge/spin correlations for both trivial quasi-helical and trivial ballistic cases are plotted in Figs.~\ref{trivial}(c,d). One can clearly conclude that while the trivial quasi-helical (spin-momentum locked in absence of non-magnetic disorder) case yields no correlations the trivial ballistic case yields a finite non-local correlation for both charge as well as spin enabling an effective distinction between the two cases. This is entirely expected since in a set-up without probe disorder as we have seen earlier for chiral as well as topological(helical) cases the non-local correlation vanishes. For the ballistic phase on the other hand because these are not edge modes they will  be backscattered in addition to being susceptible to spin flips yielding finite non-local correlations.  The results are summarized in table 5 below. 
\begin{table}[h]
\begin{center}
\begin{tabular}{|c||c|c|} 
 \hline
& Trivial quasi-helical  & Trivial Ballistic  \\ 
\hline
Non-local Charge Noise correlations & Always zero & turn completely negative  \\
\hline
Non-local Spin Noise correlations & Always zero & turn completely positive \\
\hline
\hline
\end{tabular}
\caption{Trivial quasi-helical (Spin momentum locked in absence of non-magnetic disorder) phase vs. Trivial ballistic (without spin-momentum locking) phase.}
\end{center}
\end{table}
\section{Conclusion} To conclude we establish here that we can probe helical edge modes as well as their topological or, otherwise, trivial origin via non-local HBT correlations. These correlations can be positive with topological helical QSH edge modes but will always be negative with chiral QH edge modes. Further, we show that the difference between the non-local charge and spin correlations can also distinguish between the chiral and helical edge modes. The non-local charge correlations turn completely negative for trivial quasi-helical edge modes while the non-local spin correlations turn completely positive. In Table 1 we summarize the results for the distinction between chiral and Helical edge modes while in Table II we bring out the differences between trivial and topological QSH edge modes. To end, we would like to point out that although in our work we have exclusively focused on chiral and helical edge modes and their topological origins our detection technique (Non-local HBT correlations) can be a very effective tool to probe helicity and its origin in Weyl semi metals\cite{furusaki} too.
\bibliography{qsh-noise}
\section{Acknowledgments}
This work was supported by funds from Dept. of Science and Technology (Nanomission), Govt. of India, Grant No. SR/NM/NS-1101/2011 and SCIENCE \& ENGINEERING RESEARCH BOARD, New Delhi, Govt. of India, Grant No. EMR/20l5/001836.
\section{Author contributions statement}
C.B. conceived the proposal, A.M. did the calculations on the advice of C.B., A.M. and C.B. wrote the paper, A.M. and C.B analyzed the results.   
 
\textbf{Competing financial interests} 
The author(s) declare no competing financial interests.
\textbf{Data availability statement} All data generated or analysed during this study are included in this published article.
\end{document}